\documentclass[sigconf, nonacm]{acmart}

\usepackage{balance} 
\usepackage{booktabs} 
\usepackage{amsmath}
\usepackage{graphicx}
\usepackage[linesnumbered,ruled,vlined]{algorithm2e}
\usepackage{epstopdf}
\usepackage[normalem]{ulem}
\usepackage{url}
\usepackage{enumerate}
\usepackage[labelfont=bf]{caption}
\usepackage{xcolor}
\usepackage{color, colortbl}
\usepackage{tabularray}
\usepackage{bm}

\usepackage{pifont}
\usepackage{makecell}
\usepackage{array}
\newcommand{\sys}{SELCC}

\newcommand{\myline}[1]{{\medskip\noindent\textbf{#1.}}}

\newcommand{\sect}{Sec.}
\newcommand{\fig}{Fig.}

\usepackage{indentfirst}

\usepackage[many]{tcolorbox} 
\usepackage{setspace}               
\usepackage{multicol}  

\newtcolorbox{mycodebox}{
    standard jigsaw,
    opacityback=0,  
    boxrule = 0.0pt,
    left=0pt,
    right=0pt,
    top=0pt,
    bottom=0pt,
    opacityframe=1, 
    fontupper=\footnotesize,
    sharp corners
}
\newcommand\vldbdoi{XX.XX/XXX.XX}

\newcommand\vldbavailabilityurl{URL_TO_YOUR_ARTIFACTS}
 
\definecolor{darkgreen}{rgb}{0.0, 0.5, 0.0}
\begin{document}
\pagestyle{plain} 

\title{Cache Coherence Over Disaggregated Memory}
\author{Ruihong Wang}
\affiliation{%
  \institution{Purdue University}
}
\email{wang4996@purdue.edu}

\author{Jianguo Wang}
\affiliation{%
  \institution{Purdue University}
}
\email{csjgwang@purdue.edu}

\author{Walid G. Aref}
\affiliation{%
  \institution{Purdue University}
}
\email{aref@purdue.edu}




\begin{abstract}
\noindent Disaggregating memory from compute offers the opportunity to better utilize  stranded memory in cloud data centers. It is important to cache data in the compute nodes and maintain cache coherence across  multiple compute nodes. However, the limited computing power on disaggregated memory servers makes traditional cache coherence protocols suboptimal, particularly in the case of stranded memory. This paper introduces SELCC; a Shared-Exclusive Latch Cache Coherence protocol that maintains cache coherence without imposing any computational burden on the remote memory side. It aligns the state machine of the shared-exclusive latch protocol with the MSI protocol 
, thereby ensuring both atomicity of data access and cache coherence \textcolor{black}{with sequential consistency.} SELCC embeds cache-ownership metadata directly into the RDMA latch word, enabling efficient cache ownership management via RDMA atomic operations. SELCC can serve as an abstraction layer over disaggregated memory with APIs that resemble main-memory accesses.  A concurrent B-tree and three transaction concurrency control algorithms are realized using SELCC's abstraction layer. Experimental results show that SELCC significantly outperforms Remote-Procedure-Call-based protocols for cache coherence under limited remote computing power. Applications on SELCC achieve comparable or superior performance over disaggregated memory compared to competitors.

\end{abstract}

\maketitle



\section{Introduction}
\label{sec:intro}
Memory disaggregation has emerged as a significant trend in cloud databases in both academia~\cite{ZhangCIDR20,ZhangVLDB20, Farview22, Wang023DDS,lu2024dex,ZuoSYZ021, DisArt23,DSMDB2022} and industry~\cite{PolarDBServerless21,Disaggregation21,PolarDBMP}. 
An important motivation behind disaggregated memory is to utilize the substantial amounts of stranded memory~\cite{Redy22,gu2017efficient,RaoP16,GrandlAKRA14,VermaPKOTW15} in cloud data centers. Stranded memory refers to memory that is inaccessible 
due to all the available cores being allocated to virtual machines~\cite{Redy22}. Memory disaggregation addresses this issue by accessing the stranded memory via high-speed networks, physically decoupling the memory resources from compute servers. By establishing disaggregated memory over stranded memory, cloud providers can significantly enhance memory utilization and reduce the total cost of ownership (TCO). 

\myline{One-Sided RDMA: Key to Efficient Memory Disaggregation}
The unique feature of disaggregated memory is that memory nodes 
have 
\textit{very limited} computing power. This constraint necessitates a more efficient data access method that is provided by Remote Direct Memory Access technology (RDMA, for short), particularly one-sided RDMA operations.
One-sided RDMA allows data transfer to fully bypass the CPU on remote memory, achieving low latency and minimal use of remote computing resources. 
In contrast,  traditional RPC-based access schemes become inefficient for disaggregated memory, especially when the computing power on memory nodes 
is
limited or, 
at times, 
nonexistent.

%

\myline{The Cache Coherence Problem over Disaggregated Memory}
Memory disaggregation enables sharing main memory among multiple compute nodes.
This advancement drives the next generation of multi-primary architectures~\cite{PolarDBMP, GaussDBMP} that resolve conflicts among multiple writers through disaggregated memory. The key challenge in designing multi-primary systems over disaggregated memory is \textit{maintaining  cache coherence between the compute nodes}. 
Given that RDMA latency is approximately 10 times slower than main-memory access, 
compute-side caching is necessary as it can effectively reduce these round trips by leveraging locality. However, multiple copies of data across compute nodes introduce consistency challenges, necessitating a robust software-level cache-coherence protocol to ensure data integrity.



\begin{figure*}[htbp]
\centering
\includegraphics[width=0.99\textwidth]{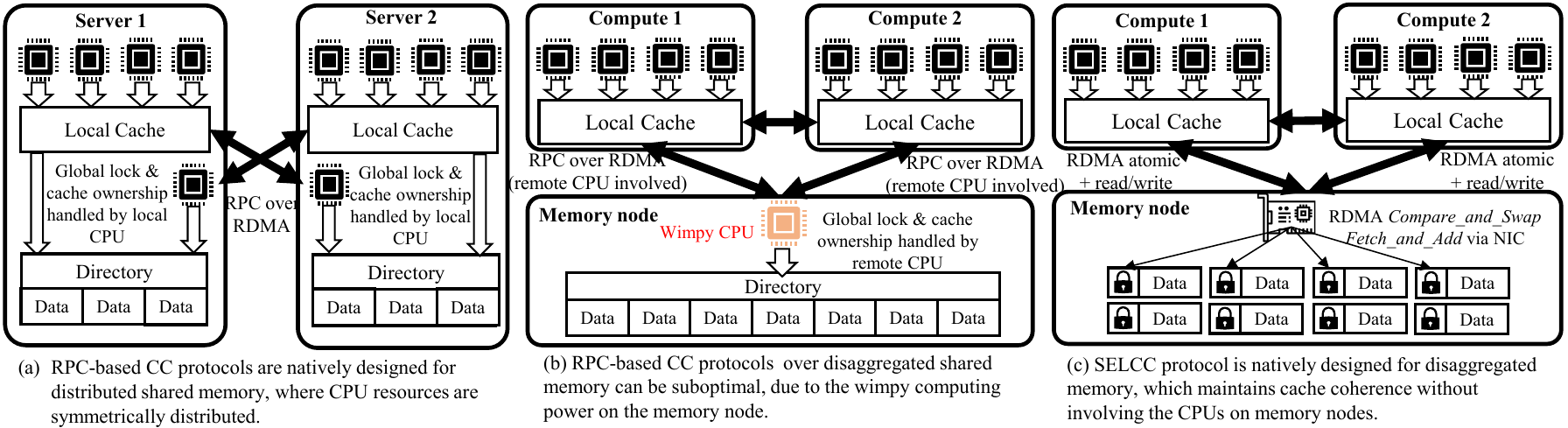}
\vspace{-0.3cm}
\caption{RPC-based cache coherence protocols vs. SELCC protocol}\label{fig:Intro}
\vspace{-0.5cm}
\end{figure*}

\myline{Limitations of Existing Cache-Coherence Solutions}
Existing cache-coherence protocols over RDMA, 
e.g., GAM~\cite{GAM18}, ScaleStore~\cite{ScaleStore22,ScaleStoreCIDR}, and the solutions  in  PolarDB MP~\cite{PolarDBMP} and GaussDB~\cite{GaussDBMP} are all RPC-based protocols, which rely on the computing resources in memory nodes. As 
in \fig~\ref{fig:Intro}a, these protocols 
have been
designed for distributed shared-memory systems, where compute and memory resources are co-located (i.e., not disaggregated). In 
these 
systems, a server can utilize abundant computing power to process cache ownership management and resolve access conflicts. However, these protocols become suboptimal when applied to disaggregated memory, particularly when the memory pool is established over stranded memory.  RPC requests  suffer due to the  scarce computing power in memory nodes (\fig~\ref{fig:Intro}b). Thus, there is pressing need for native cache coherence  over disaggregated memory that bypasses remote memory CPU when resolving conflicts.

\myline{Challenges} 
This paper investigates \textit{how to maintain cache coherence among multiple compute nodes while adhering to the one-sided access scheme between compute nodes and the disaggregated memory.}
There are many challenges: (1)~
Managing cache ownership distributively via one-sided RDMA is extremely challenging and is fundamentally different from RPC-based protocols. Simply leveraging RDMA read and write to manage the cache directory is inefficient due to the introduced RDMA round trips. 
(2)~Optimizing the protocol to minimize RDMA round-trips and bandwidth consumption is non-trivial. (3)~Maintaining fair data access among multiple compute nodes is complicated, especially given that the new protocol totally bypasses the processing of remote memory.

\myline{Our Approach}
This paper presents the Shared-Exclusive Latch-based Cache-Coherence protocol (SELCC), a new protocol for the cache coherence problem over disaggregated memory. By introducing lazy latch-release and invalidation messages, the one-sided RDMA shared-exclusive latch protocol can be upgraded to address the cache-coherence problem with \textcolor{black}{\textit{sequential consistency}}. It embeds the cache ownership information into RDMA latch words, allowing both the latch and cache ownership to be managed within a single RDMA atomic operation. 
To optimize performance, SELCC prioritizes  local ownership handover over  global ownership handover,
employs batched processing for cache eviction, and minimizes the bandwidth for dirty data flush back. Additionally, SELCC enhances the fairness by incorporating priority into invalidation messages. In \fig~\ref{fig:Intro}c, SELCC operates without involving remote memory computing power, making it particularly suited for memory disaggregation. 
Note that in scenarios where remote memory has computing resources, they can be used for other functionalities (rather than handling cache coherence), such as operator pushdown, which can result in significant performance gains~\cite{lu2024dex,dLSMICDE23}.

\myline{Contributions} 
The contributions of this paper are as follows.
(1)~We introduce SELCC, a new one-sided RDMA latch protocol that supports  RDMA access atomicity and cache coherence across multiple compute nodes (\sect~\ref{sec:overview}--\ref{sec:InsOpts}).
Compared to other cache-coherence protocols,  
SELCC is agnostic to whether or not CPU compute power is available at the disaggregated memory side. (2)~We present an API for SELCC that serves as an
abstraction layer over disaggregated memory. We demonstrate the usefulness of this API by realizing a concurrent B-tree 
and
three transaction concurrency control algorithms (\sect~\ref{sec:application}). 
 The proposed API simplifies realizing databases over disaggregated memory. While prior research works optimize indexes and transaction engines for disaggregated memory, these efforts have been studied independently. Integrating them into a single unified database is difficult due to their differing approaches to data synchronization. With SELCC, we can build indexes and transaction management directly on the same SELCC layer without 
 worrying about
 cache coherence. 

\myline{Open-source} SELCC is available at \textcolor{black}{\url{https://github.com/ruihong123/SELCC}} (around 29,200 LOC).

\section{background}\label{sec:background}
\noindent
{\bf RDMA Technology. } Remote Direct Memory Access (RDMA) is a high-speed inter-memory communication method with low latency. It allows direct access to the memory of a remote node~\cite{KaliaKA16}. RDMA bypasses the host operating system when transferring data to avoid extra data copy. RDMA's kernel-bypassing and low-latency features make it applicable to high-performance data centers~\cite{IntelRSD,IBMCloud,PolarDBServerless21,Disaggregation21}.

\textit{ibverbs} is a C++ library for RDMA programming that provides low-level implementation of RDMA primitives. There are five types of primitives in \textit{ibverbs}: RDMA send, RDMA receive, RDMA write, RDMA read, and RDMA atomic~\cite{WeiD0C18,MitchellGL13}. 
RDMA write, RDMA read, and RDMA atomic are one-sided RDMA primitives that directly access the remote server's memory without involving the remote server's CPU. Two-sided RDMA primitives (including RDMA send and RDMA receive) involve both sides of the compute and memory servers. 
RDMA atomic includes two primitives: \textit{RDMA compare and swap} (\texttt{RDMA\_CAS}) and \textit{RDMA fetch and add} (\texttt{RDMA\_FAA}). These primitives ensure the atomicity of a group of operations on data of at most 8 bytes. Additionally, \texttt{RDMA\_CAS} and \texttt{RDMA\_FAA} can be leveraged to implement shared-exclusive latch over RDMA (SEL), guaranteeing atomicity among RDMA reads and writes~\cite{Tobias23}.

\myline{Cache-Coherence Protocols}
Cache coherence is a concept in multiprocessor systems ensuring that multiple data copies in various CPU caches remain consistent~\cite{culler1999MSI}. In multiprocessor systems,  consistency is ensured via hardware-level cache-coherence protocols. In disaggregated memory systems, 
hardware-level protocols are not present. Thus, a software-level cache-coherence mechanism is needed when local caches are deployed in compute nodes. 

Existing cache-coherence protocols~\cite{GAM18,ScaleStore22,amza1996treadmarks,CarterBZ91,LiH89,StetsDHHKPS97} over RDMA have been originally designed for distributed shared memory systems, where each object has a main copy stored in its home node. These protocols maintain cache-coherence using methods similar to those for multiprocessor systems, e.g., MSI, MESI, and MOESI~\cite{culler1999MSI},
and utilize state machines to manage different ownership types. 
Indexes  over disaggregated memory bypass the cache coherence problem by caching only the index metadata, e.g., B-tree internal nodes or hash directory)~\cite{DisArt23,Sherman2022,ZuoSYZ021}. This approach is effective but has limitations: (1)~These caches are typically limited in size and cannot be adjusted to match the available local memory capacity. (2)~Metadata caching is typically specific to particular data structures, limiting its generality. Thus, there is significant need for a general cache coherence protocol that eliminates the need for computing over remote memory.



\section{System Overview}
\label{sec:overview}

\begin{figure}[htbp]
\vspace{-0.4cm}
\centering
\includegraphics[width=0.38\textwidth]{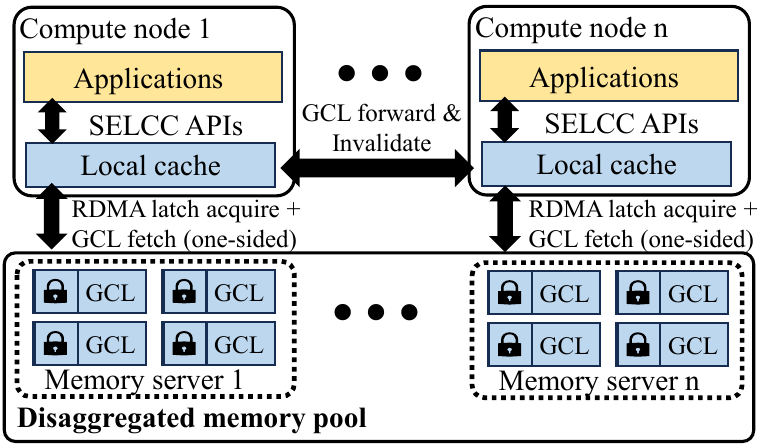}
\vspace{-0.4cm}
\caption{System overview}\label{fig:Overview}
\vspace{-0.5cm}
\end{figure}

This paper addresses optimizing DBMSs for disaggregated memory~\cite{DSMDB2022,PolarDBMP,GaussDBMP}, where compute and memory are decoupled, and multiple compute nodes share a common memory pool (\fig~\ref{fig:Overview}). 

The memory pool consists of multiple memory servers. Data  is addressed via an 8-byte global pointer {\em (NodeID, offset)}, where {\em NodeID} is the memory server identifier, and {\em offset} is the  offset within  server memory.  
In \fig~\ref{fig:Overview}, the disaggregated memory space is split into blocks of configurable sizes, referred to as Global Cache Lines (GCLs). GCL is the main data manipulation unit between the compute and memory nodes, and has 3 components: A one-sided global latch word (8 bytes), a user-defined header, and the data region. The global latch word is a crucial and ensures one-sided RDMA access atomicity and cache coherence. The user-defined header is  application-specific, similar to page headers in traditional disk-based databases. Finally, the data region stores data objects, e.g., tuples for data tables and key-value pairs for indexes.

When a GCL is accessed, the system first checks  local cache. If the GCL is not found or its ownership state is incorrect, the system attempts to acquire the corresponding RDMA latch and fetch the latest GCL using \texttt{RDMA\_CAS} and \texttt{RDMA\_Read} within a single 
RDMA round trip. If the lock acquisition fails, which indicates a conflicting copy on another compute node, invalidation messages will be initiated to force the current owner to transfer  global ownership as well as the latest copy (More on this in Sec.~\ref{sec:conflictpath}). 
SELCC exposes a simple API to  applications that allows users to bypass the complexities of RDMA programming.  Many data structures and algorithms for monolithic servers can be migrated onto SELCC seamlessly (\sect~\ref{sec:application}), as the RDMA access atomicity and cache coherence problem has already been resolved underneath this API.

\begin{figure*}[htbp!]
\centering
\includegraphics[width=0.9\textwidth]{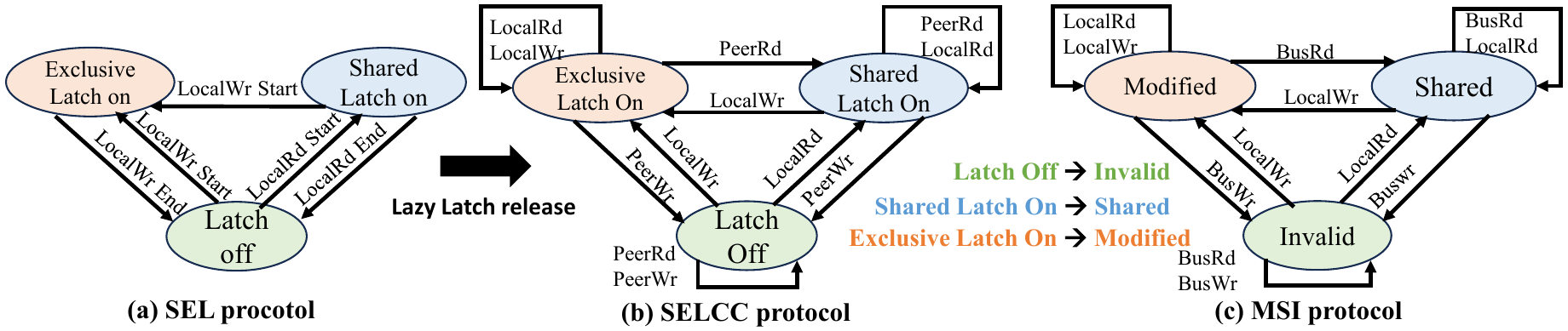}
\vspace{-0.4cm}
\caption{State machines for SELCC, SEL, and MSI protocols }\label{fig:StateMachine}
\vspace{-0.1cm}
\end{figure*}

\section{The SELCC Protocol}
\label{sec:SELCC}
\begin{sloppypar}
We introduce the SELCC Protocol; 
\underline{\bf S}hared-\underline{\bf E}xclusive \underline{\bf L}atch-based \underline{\bf C}ache \underline{\bf C}oherence protocol (SELCC). It guarantees cache coherence and atomicity for concurrent RDMA reads and writes. 
We follow the established design practice in \cite{Tobias23,KaliaKA16} to ensure both correctness and efficiency.

\end{sloppypar}
\subsection{Protocol Overview}
\label{sec:SELCCOV}


\noindent{\bf Main Idea. } The SELCC protocol draws inspiration from the one-sided Shared-Exclusive Latch protocol (SEL)~\cite{Tobias23} and the Modified, Shared, Invalid  protocol (MSI)~\cite{culler1999MSI}.
SEL uses RDMA atomic operations to ensure the atomicity of RDMA accesses, enabling concurrent RDMA reads through the use of shared state. MSI  is a widely adopted cache coherence protocol  in multiprocessor systems. It manages the cache ownership states (Modified, Shared, and Invalid) via a state machine for consistent  data reads and writes (\fig~\ref{fig:StateMachine}c). 



Interestingly, we observe that the semantic meanings of  MSI states are analogous to those of  SEL. In SEL, the \textit{Exclusive} state indicates a locally modified copy, the \textit{Shared} state indicates a locally shared copy, and the \textit{Latch Off} states indicates an invalid local copy. However, the conditions triggering state transitions differ significantly. In  SEL, compute nodes eagerly release an RDMA latch once local access is complete, leading to immediate invalidation of data copies (\fig~\ref{fig:StateMachine}a). 
In contrast, MSI has a lazy invalidation strategy, where cache states are invalidated only upon receiving bus signals from other processors. \textbf{By mapping cache states to latch states and synchronizing  SEL's state machine with that of  MSI's, we can resolve the cache-coherence problem.}

To align the state machines, SELCC introduces two key mechanisms: lazy latch-release and invalidation messages (PeerRd, PeerWr), as in \fig~\ref{fig:StateMachine}b. When a compute node  acquires the latch, the fetched copy is stored in local cache. Unlike traditional latch mechanisms, SELCC defers the release of the latch until either another compute node accesses the same GCL or the corresponding cache frame is evicted. This deferred latch-release allows SELCC's state machine to align seamlessly with the MSI, as in \fig~\ref{fig:StateMachine}b and~\ref{fig:StateMachine}c. When a compute node fails to acquire the global latch, an invalidation message is issued, prompting the current owners to 
release 
the latch. 

\begin{figure*}[htbp]
\centering
\includegraphics[width=0.90\textwidth]{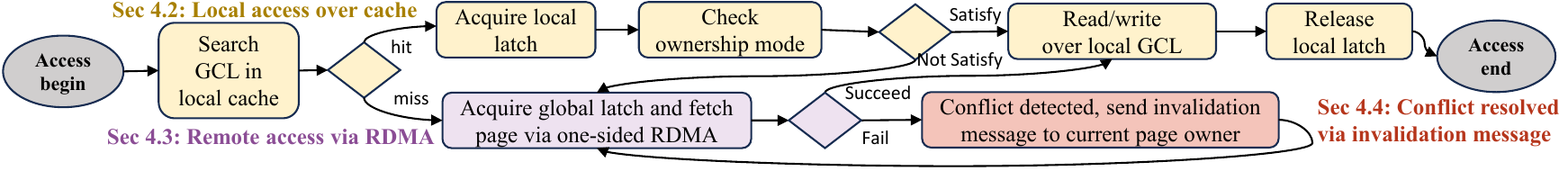}
\vspace{-0.4cm}
\caption{Protocol flow}\label{fig:protocol}
\vspace{-.4cm}
\end{figure*}

\noindent\textbf{SELCC Flow.} Data access in SELCC is divided into three phases  (\fig~\ref{fig:protocol}). (1)~The accessing thread searches  local cache for the target Global Cache Line (GCL), leveraging access locality to minimize RDMA round trips (\sect~\ref{sec:localpath}). 
(2)~If no valid cache frame is found, the thread employs one-sided RDMA to retrieve the latest copy from disaggregated memory (\sect~\ref{sec:remotepath}). 
(3)~If conflicting cache copies are found in other compute nodes, e.g., an exclusive copy in another node, the conflict is resolved by invalidation messages (\sect~\ref{sec:conflictpath}). 

\noindent\textcolor{black}{\textbf{Challenges.} The key challenge lies in efficiently managing ownership using one-sided RDMA while ensuring it incurs no additional round trips compared to RPC-based solutions. Each phase of the process presents specific challenges: (1)~How to effectively leverage metadata in local cache frames to minimize remote accesses? (2)~How to efficiently acquire ownership in remote memory through one-sided RDMA operations? (3)~How to correctly and efficiently transfer ownership and the latest GCL copies across compute nodes? This challenge becomes particularly complex in the presence of varying conflict scenarios, each of which requires a tailored design.} 

\subsection{Local GCL Access via Cache}
\label{sec:localpath}

The local cache frame not only stores a copy of the Global Cache Line (GCL) but also maintains an ownership memo and a local shared-exclusive latch. The ownership memo records the acquired global ownership (Shared, Modified, or Invalid) on this data copy. The later accesses can verify whether they are permitted without involving any RDMA operation.
The local latch synchronizes concurrent accesses to the cache frame between local accessing threads and invalidation message handlers.

As in \fig~\ref{fig:protocol}, when accessing a GCL, the thread  searches  local cache for the target GCL. In case of a cache miss, the thread fetches the data and acquires ownership from the disaggregated memory. For a cache hit, the thread  acquires the local latch, and then checks against the ownership memo. If the ownership satisfies the current access type, the thread can directly access  data via local cache. Else, RDMA operations are invoked to acquire the appropriate ownership (\sect~\ref{sec:RemoteAcquire}) and the latest GCL copy. Ownership verification rules are given in the table below, where rows represent the type of local access, and columns represent the ownership states in the memo.
\begin{table}[h]
\vspace{-0.5cm}
\resizebox{\columnwidth}{!}{%
\begin{tabular}{l|llll}
\cline{1-4}
\textbf{} & \textbf{Modified} & \textbf{Shared}                     & \textbf{Invalid / cache miss}                     &  \\ \cline{1-4}
\textbf{Reader}    & \textcolor{darkgreen}{\checkmark}        & \textcolor{darkgreen}{\checkmark}                     & \textcolor{purple}{Acquire Shared ownership}   &  \\ \cline{1-4}
\textbf{Writer}    & \textcolor{darkgreen}{\checkmark}        & \textcolor{purple}{Ownership upgrade} & \textcolor{purple}{Acquire Modified ownership}&  \\ \cline{1-4}
\end{tabular}%
}
\label{tab:my-table}
\end{table}
Once the corresponding ownership is acquired and  local cache access is completed, the accessing thread releases the local latch, while leaving the global ownership memo unchanged.

\subsection{Remote GCL Access via One-Sided RDMA}
\label{sec:remotepath}
When no valid copy of the target GCL is present in  local cache, the accessing thread must retrieve the latest GCL via one-sided RDMA and update global ownership in the disaggregated memory. In RPC-based protocols, there are message handling threads in the disaggregated memory, managing the cache ownership directory of GCLs. However, maintaining this ownership directory via one-sided RDMA is challenging, particularly when no additional RDMA round-trip is expected compared to the RPC-based solution.

\subsubsection{\textbf{Embeded Ownership in RDMA Latch Word}}
\begin{figure}[tbp]
\centering
\includegraphics[width=0.30\textwidth]{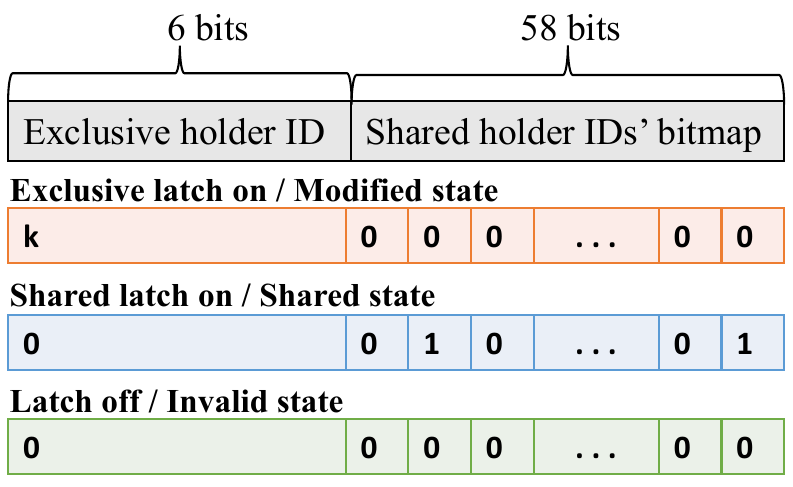}
\vspace{-0.2cm}
\caption{RDMA latch words in SELCC}
\vspace{-0.56cm}
\label{fig:latchword}
\end{figure}

In SELCC, we propose 
to embed  cache ownership within the RDMA latch words, allowing both cache ownership and RDMA latch  to be managed via one RDMA atomic operation.
As in \fig~\ref{fig:latchword}, a 64-bit latch word is the maximum data length supported by RDMA atomic operations. We divide these 64 bits into 2 parts: (1)~An exclusive latch holder's ID (6 bits), and (2)~A reader holders' ID bitmap (58 bits).\footnote{This protocol can support up to 58 compute nodes. With multi-cores  on each compute node, a system using SELCC can support thousands of cores.}

This new structure allows the RDMA latch to simultaneously track both shared and exclusive latch holder IDs. For instance, if a node with ID $A$ 
holds a modified copy of a GCL, say $g$, the latch word associated with $g$ is represented as $(A, 000...0)$, where $A$ is the exclusive holder ID and the reader bitmap is cleared. When two nodes, say $A$ and $B$, concurrently hold shared copies of $g$, the latch word is represented by $(0, (1<<A) | (1<<B))$, where no exclusive holder exists, and the bit positions for $A$ and  $B$ are set to 1 in the reader bitmap. If there is no valid cached copy in the compute nodes, the RDMA latch is set to $(0, 00..0)$.
This approach has two benefits: \textbf{(1)~No additional RDMA round trips are needed to maintain the cache directory,  (2)~Atomicity of directory changes is ensured}.  
If lock acquisition fails, a compute node  gets the latest cache holders' IDs via the RDMA atomic operation's return value, enabling the determination of invalidation message recipients. 
\subsubsection{\textbf{Acquiring Ownership via One-Sided RDMA}}
\label{sec:RemoteAcquire}
With embedded ownership in an RDMA latch word, acquiring ownership globally is equivalent to acquiring the corresponding RDMA latch from disaggregated memory. Based on the type of access and the current ownership mode in the cache frame, there are three  types of ownership requests implemented via one-sided RDMA.

\noindent\textbf{Acquiring ``Modified" Ownership. }
To acquire {\em Modified} ownership, a writer atomically compares the latch word with $(0, 00..0)$, and swaps it with $(NodeID, 00..0)$. If the \texttt{RDMA\_CAS} fails, the prior latch word is returned to the compute node that parses the shared/ex\-clu\-sive latch holder IDs and sends them invalidation messages. 

\noindent\textbf{Acquiring ``Shared" Ownership.}
To acquire shared ownership, a reader atomically fetches the latch word and sets its own position in the bitmap using \texttt{RDMA\_FAA} (with the add operand set to $(1 << NodeID)$). The reader  checks the return value of \texttt{RDMA\_FAA} to determine if a writer is currently holding the latch. If so, the acquisition fails, and the reader resets its bit in the bitmap using another \texttt{RDMA\_FAA}.  Then it  sends an invalidation message based on the holder ID of the exclusive latch. 

\begin{sloppypar}
\noindent\textbf{Upgrading ``Shared" to ``Modified" Ownership.}
When a writer finds that the target GCL is cached in ``Shared" state, the compute node needs to upgrade the global latch from shared to exclusive. First, the compute node  attempts to compare and swap the global latch words from $(0, 1 << NodeID)$ to $(NodeID, 00..0)$. If two nodes simultaneously upgrade the same global latch, both nodes will continuously fail at the CAS operation. This issue is analogous to the deadlock problem for lock upgrade within a single machine, and is resolved using a typical fallback approach.\footnote{Note that global lock upgrading is always triggered implicitly, and the atomic ownership/latch upgrade for the user is not necessary.} After several failed attempts to upgrade the latch, SELCC resorts to releasing the shared latch \textcolor{black}{and then} acquiring the exclusive latch.


\begin{figure}[htbp]
\vspace{-0.3cm}
\centering
\includegraphics[width=0.40\textwidth]{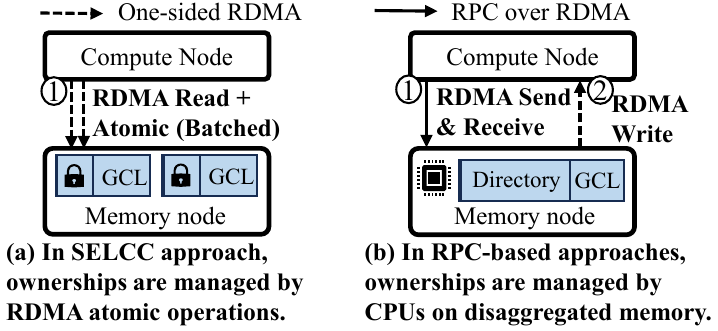}
\vspace{-0.2cm}
\caption{Different approaches for remote access}
\vspace{-0.4cm}
\label{fig:RemoteRT}
\end{figure}

As  in \fig~\ref{fig:RemoteRT}a, SELCC issues RDMA read and atomic operations to atomically retrieve the latest data copy and modify the latch state (ownership). 
Since RDMA read and atomic operations are conducted in batches, only one  RDMA round-trip is needed for this combined operation. Contrasting to the RPC-based solution (\fig~\ref{fig:RemoteRT}b), SELCC 
reduces latency, as RPC-based protocols typically require two RDMA operations issued separately on both sides.
\end{sloppypar}

\subsection{Resolving Conflicted GCL Accesses}
\label{sec:conflictpath}
When a thread fails to acquire the global latch for a GCL via RDMA, it issues an invalidation message (\sect~\ref{sec:InvalidationM}) to prompt the current owner to either relinquish or transfer its ownership, as well as the data copy, and, if necessary, flush back the dirty GCL.
There are three types of conflict scenarios:
(1)~A writer on the sender side invalidates a modified copy on the receiver side.
(2)~A reader on the sender side invalidates a modified copy on the receiver side.
(3)~A writer on the sender side invalidates one or more shared copies on the receiver sides. Each scenario must be handled differently.
\subsubsection{\textbf{Case 1: A Writer on the Sender Side Invalidates a ``Modified" Copy on the Receiver Side}}
\label{sec:PageForward}
Refer to~\fig~\ref{fig:WRInvM}  for illustration.
Node $A$ attempts to \textbf{write} a GCL while Node $B$ holds a copy in the ``\textbf{Modified}" state.
To resolve this conflict, the naive approach is that the message receivers simply release the global latch and actively flush back the dirty data to disaggregated memory. Meanwhile, the message senders repeatedly attempt to acquire the latch and retrieve the latest data from disaggregated memory (\fig~\ref{fig:WRInvM}a), following the procedure outlined in Section~\ref{sec:RemoteAcquire}.
However, this naive approach is inefficient for two reasons: (1)~The retry mechanism consumes a large amount of RDMA bandwidth between compute and memory nodes. (2)~Compared to RPC-based solutions (\fig~\ref{fig:WRInvM}c), it takes four RDMA round-trips, higher than three in RPC-based solutions. To optimize it, we introduce two new techniques: \textit{Global Ownership Handover} and \textit{GCL Forwarding}.


\textit{Global Ownership Handover} allows the receiver of an invalidation message to transfer ownership directly to the requester through one atomic \texttt{RDMA\_FAA} operation, rather than requiring the sender to repeatedly attempt latch acquisition (\fig~\ref{fig:WRInvM}b)\textcolor{black}{~\cite{JohnsonPA09,Joshi91}}. 
For example, if Node~$A$ (Writer) issues an invalidation message to Node~$B$ (``Modified" state), where $A$ and $B$ are the Node IDs, ownership can be atomically transferred from Node~$B$ to Node~$A$ by adding the value $(A - B, 00..0)$ onto the latch word via \texttt{RDMA\_FAA}. 

\textit{GCL Forwarding} enables a message sender to retrieve the latest GCL copy directly from the message receiver, bypassing the need to fetch it from disaggregated memory. The sender includes its local buffer address in the invalidation message, allowing the receiver to write back the latest GCL via an RDMA write. Additionally, if GCL ownership is transferred between nodes as a ``Modified" copy, the dirty GCL does not need to be flushed back to disaggregated memory, as the following owner always acquires the latest copy from the last exclusive owner in a compute node.

As in \fig~\ref{fig:WRInvM}b, SELCC requires only three RDMA round trips to handle 
this
cases, which is the same as RPC-based solutions. \textcolor{black}{Furthermore, SELCC fully bypasses  remote processing, making it  effective over even stranded memory.} Notably, the victim of an invalidation message can immediately forward the dirty GCL to the sender's local buffer without waiting for the ownership handover round-trip (\texttt{RDMA\_FAA}). Although this design may result in 
outdated
invalidation messages being sent by compute nodes, outdated messages can simply be discarded upon detection.

\subsubsection{\textbf{Case 2: A Reader on the Sender Side Invalidates a ``Modified" Copy on the Receiver Side}}
\begin{figure}[htbp]
\centering
\vspace{-0.2cm}
\includegraphics[width=0.5\textwidth]{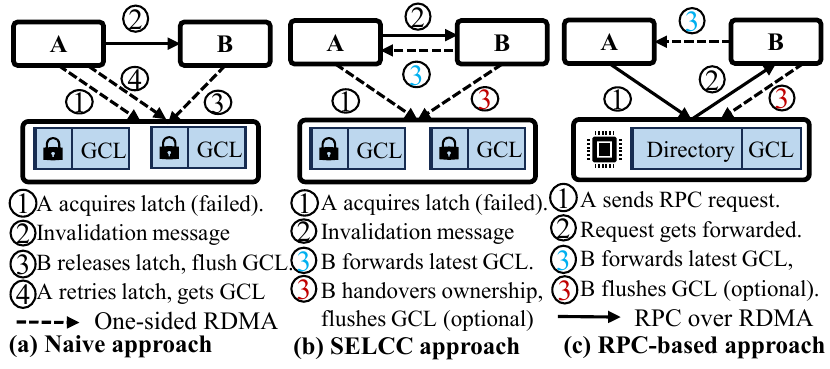}
\vspace{-0.8cm}
\caption{Different approaches to resolving conflict when a writer/reader invalidates a copy in ``Modified" state.}
\vspace{-0.6cm}
\label{fig:WRInvM}
\end{figure}

Refer to 
\fig~\ref{fig:WRInvM} for illustration. Node $A$ attempts to \textbf{read} a GCL while Node $B$ holds a copy in the ``\textbf{Modified}" state.
This scenario is generally handled in a manner similar to Case 1, with minor adjustments for global ownership handover and GCL forwarding. As  in \fig~\ref{fig:WRInvM}b, when the current owner in the ``Modified" state receives an invalidation message, it atomically modifies the latch word by adding $(-B, (1 << A)|(1 << B))$ via \texttt{RDMA\_FAA}. This operation clears the modified ownership for Node $B$ and assigns shared ownership to both Nodes $A$ and $B$. \textcolor{black}{Additionally, in the same RDMA round trip, the latest GCL is flushed back to the disaggregated memory together (unlike Case 1), ensuring that future concurrent readers can acquire the latest ``Shared" copy from the disaggregated memory rather than fetch it from another compute node.} Finally, the latest GCL is forwarded to the sender. 




\subsubsection{\textbf{Case 3: A Writer on the Sender Side Invalidates One or More ``Shared" Copies on the Receiver Sides}}
 \label{sec:InvCase3}
\begin{figure}[tbp]
\vspace{-0.3cm}
\centering
\includegraphics[width=0.46\textwidth]{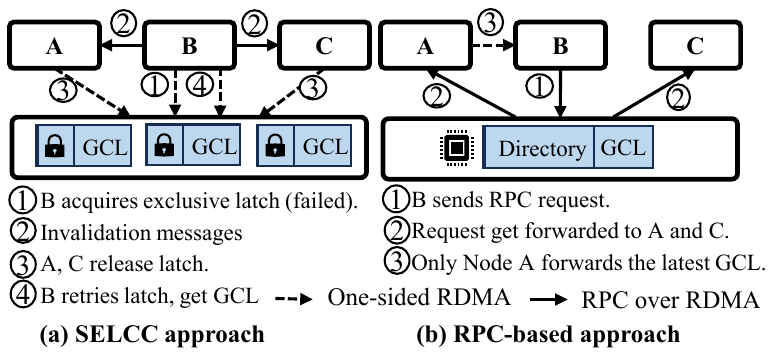}
\vspace{-0.2cm}
\caption{Different approaches to resolving conflict when a writer invalidates one/many ``Shared" copies.}
\vspace{-0.5cm}
\label{fig:WInvR}
\end{figure}

Refer to 
\fig~\ref{fig:WInvR} for illustration. Node $B$ attempts to \textbf{write} a GCL, while Nodes $A$ and $C$ hold copies in the ``\textbf{Shared}" state.
Unlike the previous case, GCL forwarding and ownership handover cannot be applied here, as atomic ownership transfer among multiple nodes with shared copies is impossible. For example (\fig~\ref{fig:WInvR}a), consider a scenario where Node $B$ (writer) identifies shared copies on Nodes $A$ and $C$. $A$ and $C$ cannot collectively transfer the ``Shared" ownership to $B$ in the latch word atomically. Thus, one node must take the responsibility as a leader to manage ownership on behalf of the others. Even if Node $A$ is designated as the leader, it cannot determine whether another node, e.g., $D$, has successfully acquired shared ownership during the message transmission. In these cases,  $D$'s ``Shared" ownership is overlooked by $A$ (the leader), potentially leading to corruption of the RDMA latch word.

As a result, our design in SELCC is as follows. 
Refer to \fig~\ref{fig:WInvR}. The invalidation message prompts each victim node to release its shared lock while the sender continues attempting to acquire the exclusive latch. Compared to an RPC-based solution (\fig~\ref{fig:WInvR}b), 
SELCC
may have one extra RDMA round-trip, but given that it fully bypasses the need for remote compute power, this trade-off is acceptable.

\section{Optimizations}
\label{sec:InsOpts}
We implement  SELCC  into the compute-side cache over  disaggregated memory. 
The compute-side cache is a lightweight hash table with LRU replacement policy and is sharded to support high concurrency. While  SELCC  resolves the cache-coherence problem, several instantiation challenges remain unaddressed: (1)~Efficiently implementing invalidation messages across compute nodes, (2)~Optimizing cache eviction to minimize its impact on read and write operations, and (3)~Avoiding latch starvation to maintain fairness among the compute nodes.


\begin{figure}[htbp]
\vspace{-0.1cm}
\centering
\includegraphics[width=0.45\textwidth]{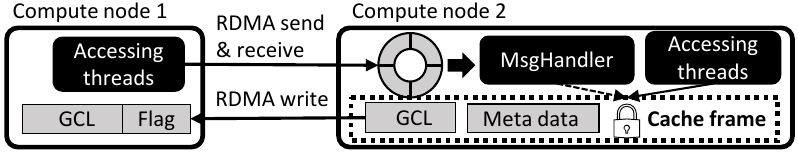}
\vspace{-0.2cm}
\caption{Invalidation message}
\vspace{-0.2cm}
\label{fig:InvMsg}
\end{figure}
\subsection{Efficient Invalidation Messages}
\label{sec:InvalidationM}
Invalidation messages, implemented via RPC over RDMA, play a critical role in coordinating global cache-line (GCL) ownership in SELCC. 
Invalidation messages contain key information, 
e.g.,
the global address of the target GCL and the case  of invalidation (\sect~\ref{sec:conflictpath}).  
As  in \fig~\ref{fig:InvMsg}, the RPC request is sent via \texttt{RDMA\_send}, to the receiver side maintaining a ring buffer to capture incoming messages via RDMA receive. 
The background message handlers in the receiver node 
process the invalidation messages by releasing/handing over the global latch. After processing, the handler sends an acknowledgment along with the latest GCL copy to a local buffer on the sender's side via RDMA write. 
\textcolor{black}{
Importantly, in SELCC, RPC is used only between compute nodes, while the communication between compute and memory layers are strictly one-sided. SELCC adheres to the compute-free design principle for memory nodes, distinguishing it fundamentally from RPC-based protocols.
}

The efficiency and robustness of  SELCC  are supported by two key design choices: (1)~Prioritizing local accesses over invalidation message processing, and (2)~Implementing a message drop-and-resend mechanism to handle scenarios where  invalidation cannot be processed immediately.

\subsubsection{\textbf{Lower Priority for Processing Invalidation Messages}}
\label{sec:LowPrior}
\begin{figure}[htbp]
\vspace{-0.4cm}
\centering
\includegraphics[width=0.70\columnwidth]{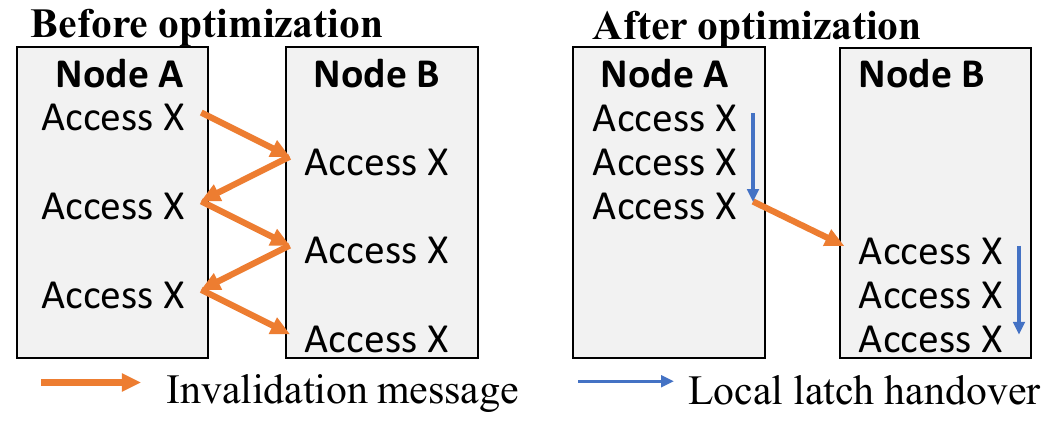}
\vspace{-0.4cm}
\caption{SELCC prioritizes local ownership transfer over global ownership transfer.}
\vspace{-0.5cm}
\label{fig:LVSG}
\end{figure}

In scenarios where invalidation messages are handled concurrently with front-end accessing threads, synchronization is required for message handlers to manage access to cache frame metadata. A straightforward approach is to acquire the local exclusive latch before processing the invalidation messages. However, this can lead to two problems: (1) the message handler can get blocked if the local accessing thread holds the local latch for an extended period, preventing the processing of subsequent invalidation messages, and (2) such a design assigns equal priority to local accessing threads and global accessing threads for handing over the data ownership. In workloads with high contention, as illustrated in \fig~\ref{fig:LVSG} left, this can lead to frequent ownership transfers among compute nodes, significantly increasing read/write latency and generating a large volume of invalidation messages.

To address these issues, SELCC adopts a design that assigns lower priority to invalidation message handlers compared to local accessing threads. This is achieved through the use of the \texttt{try\_lock}, which attempts to acquire the local latch without blocking the handler. If \texttt{try\_lock} fails, the handler either defers or drops the message and proceeds to the next one. This approach ensures that global ownership transfers have lower priority than local ownership transfers within the same node, thus reducing latency and conserving RDMA bandwidth, as shown in \fig~\ref{fig:LVSG}.
However, prioritizing local access over global access can potentially prevent invalidation messages from taking effect under highly skewed workloads, potentially leading to global starvation on other compute nodes. The solution to this starvation problem is presented in \sect~\ref{sec:starvation}.

\subsubsection{\textbf{Message Dropping and Resending Mechanism}}
Another important design is the message drop and resending mechanism. Invalidation messages can be dropped by RPC handlers under three specific conditions: (1) the cached entry has already been invalidated by other compute nodes, (2) the target cache line has been evicted, or (3) the target GCL is currently being accessed by a local thread. When a message is dropped, the receiver writes a "dropped" flag to the end of the reply buffer in the sender side via RDMA (\fig~\ref{fig:InvMsg}). Upon receiving this flag, the message sender retries the global latch to update its view of valid cache copies and adjusts the targets of subsequent invalidation messages accordingly.
To prevent network saturation due to excessive message resending, the protocol enforces a time interval $T = \frac{C\times RTT}{N_{r}}$
between each resend, where $C$ is an empirical constant, $RTT$ is the round-trip time for RDMA atomic operations, and $N_{r}$ is the number of global latch retries. This interval also plays a crucial role in maintaining fairness among compute nodes, as will be discussed in \sect~\ref{sec:starvation}

\subsection{Efficient Cache Eviction }
Cache eviction is also an important part of SELCC. 
In SELCC, cache eviction is managed by background threads, which monitor the length of the GCL free lists and initiate eviction when the list size falls below a predefined threshold. Importantly, the cache replacement policy is independent of the cache coherence protocol, allowing SELCC to be integrated with any replacement strategy. We employ the Least Recently Used (LRU) replacement policy to select cache frames for eviction. 

The cache eviction involves two main steps: releasing ownership and, if necessary, flushing back dirty data. Releasing ownership is equivalent to releasing the global latch. When releasing an exclusive latch, the compute node atomically fetches and decrements the latch word by $(NodeID, 000...0)$, while any dirty GCLs are flushed back using RDMA write within the same RDMA round trip. We do not adopt the method from~\cite{Tobias23} for releasing the exclusive latch via \texttt{RDMA\_CAS}, as this can lead to spurious failures due to concurrent read lock operations, potentially resulting in livelock. For releasing the shared latch, the compute node resets its corresponding bit in the bitmap using \texttt{RDMA\_FAA}.

However, having the background thread release latches and flush back dirty data for every cache frame is not efficient because: (1) each GCL eviction requires at least one RDMA round trip, and if the background thread cannot keep up with eviction requests, additional RDMA round trips may be added to the critical path of read and write over SELCC; and (2) the bandwidth consumed by flushing dirty GCLs is non-trivial. To address these challenges, we next introduce two optimizations in SELCC.

\subsubsection{\textbf{Batched Processing for GCL Eviction}}
\label{sec:batcheviction}
To enhance the efficiency of GCL eviction, we process evictions in batches. The eviction worker continuously monitors the length of the GCL free list $l_{f}$ and compares it against a predefined threshold $L$. If  $l_{f}<L$, the eviction worker selects $L - l_{f}$ victims from the back of the LRU list. 
The selected GCLs are then grouped according to their node ID, which represents the memory node they belong to. This allows all RDMA evictions within the same group to be processed in a single batched RDMA work request. Importantly, the eviction worker does not need to wait for RDMA operation completion and can immediately proceed to the next batch of victims. However, this approach presents two technical challenges: (1) The local cache frames for dirty GCL cannot be immediately reused until the RDMA operation for flushing is complete. 
(2) There is a limit on the total number of outstanding RDMA operations per queue $L_{out}$\footnote{Here, "outstanding" refers to RDMA requests that have been submitted to the queue pair but have not yet been completed. Exceeding the limit of outstanding operations can lead to errors in the queue pair, disrupting communication between two machines.}. 

To address these issues, we implement a ring buffer for each queue pair connected to remote memory. The ring buffer size is set to match the maximum number of outstanding RDMA work requests per queue pair,
$L_{out}$. Before executing RDMA operations, the eviction worker attempts to acquire an available slot from the ring buffer and copies the payload from the cache frame into this buffer slot. If no slot is available, the worker must busy-wait for the completion of previous RDMA operations. Upon receiving $n$ completion notifications from the completion queue, the ring buffer’s tail is advanced by $n$ positions.
This design ensures that the number of outstanding RDMA requests always be within the threshold  $L_{out}$, and the cache frames can be immediately recycled once the dirty GCL is copied to the ring buffer. \textcolor{black}{However, these performance gains are not achieved without cost. The introduced ring buffer incurs a space overhead that scales linearly with both the number of threads per node and the number of compute nodes in the system.}

\subsubsection{\textbf{Dynamic Dirty Boundaries for GCL Flushing }}
\label{sec:Dboundaries}
To reduce RDMA bandwidth during GCL eviction, it is essential to minimize the payload size for flushing dirty GCLs. The GCL header maintains two boundaries, $d\_lower$ and $d\_higher$, which define the address range encompassing all modifications made since the GCL was fetched into the local cache. Initially, these boundaries are both set to zero, indicating that no modifications have occurred. When the first modification over GCL comes in, the $d\_lower$ and $d\_higher$ will be initialized as the start and end addresses of this modification. Afterwards, whenever a new modification arrives, $d\_lower$ will be updated as $min(d\_lower, start)$ and $d\_lower$  will be updated as $max(d\_higher, end)$.
When flushing a GCL, the eviction thread checks the dirty boundaries and only writes back the data within the specified range, rather than the entire GCL, thereby reducing RDMA bandwidth consumption.



\subsection{Fairness}
\label{sec:starvation}
Fairness is a significant challenge for the SELCC protocol, as it is based on a shared-exclusive spinlock. \textcolor{black}{A server may experience starvation if it fails to acquire the latch repeatedly.}
Next, we present the root causes of latch starvation over SELCC and propose relevant solutions accordingly. Due to the two-level hierarchy of the system, two root causes of latch starvation can be identified, each requiring distinct resolution techniques.

\noindent\textbf{Root cause 1: asymmetric local latch acquisition.} As stated in \sect~\ref{sec:LowPrior},  
to minimize the volume of invalidation messages traffic, local front-end accessing threads have higher priority than invalidation message handlers when acquiring the local latch. 
A compute node can experience global latch starvation for a particular data object if a peer compute node with a valid copy continuously receives local access requests from multiple threads for that data object. In this scenario, the local accessing threads continuously hold the local latch, causing the invalidation message handler's \texttt{try\_Lock} requests to fail continuously, leading to global latch starvation. 

\noindent\textbf{Root cause 2: asymmetric global latch acquisition.} It is not necessary to have symmetric hardware configurations across all the compute nodes. Consequently, some compute nodes with weak CPU or network resources may experience latch starvation due to the low frequency of RDMA latch retries. Additionally, if there are continuous global read requests for a particular data object, a write request for that data object may struggle to acquire the exclusive latch because peer compute nodes continuously hold the shared latch, preventing the writer from obtaining the exclusive latch. 

\subsubsection{\textbf{Handling Local Latch Starvation}}
\label{sec:FairnessType1}
To address local latch starvation, we implement a lease mechanism that forces the compute node to release the global latch when a data object has been continuously accessed by local front-end threads for an extended period. 
To interrupt these continuous local accesses at an appropriate time, two counters, the read access counter ($R_{c}$) and the write access counter ($W_{c}$), are maintained in each cache entry. These counters are activated only when an invalidation message is dropped due to the ongoing local access and is deactivated when a thread acquires the latch without spinning, indicating that the data is no longer heavily accessed.
The counters are incremented by 1 when a local access waits for the latch. Synthetic access times for the cache entry are calculated by 
$A_{times} = \frac{R_{c}}{P} + W_{c} $, where $P$ represents the number of front-end threads on the compute node. When the synthetic access times $A_{times}$ exceed a predefined threshold $\gamma$, the local thread proactively release the global latch and reset the counters. 

\subsubsection{\textbf{Handling Global Latch Starvation}}
\label{sec:FairnessType2}

To handle global latch starvation for asymmetric hardware among compute nodes, we adopt a priority aging mechanism, originally devised to solve the starvation problem in CPU scheduling~\cite{Tanenbaum09}. 
In SELCC, each invalidation message is assigned a priority that is positively correlated to the number of retries a compute node has conducted for a particular RDMA latch. The global ownership handover mechanism introduced in \sect~\ref{sec:PageForward} takes starvation priority as a key factor in determining the next owner of the data. The exclusive latch holder, receiving invalidation messages from all conflicted servers, acts as a centralized decision-maker for global latch ownership transfer. 
During the continuous local access (Sec.~\ref{sec:FairnessType1}), the invalidation message handler keeps receiving invalidation messages from other compute nodes and memorizes those messages as well as their priority. Upon handing over the global ownership, it is deliberately transferred to the sender with the highest priority. 
Furthermore, as in \sect~\ref{sec:InvalidationM}, there is a manually injected time interval between each retry for a particular latch. This interval decreases as the priority of latch acquisition increases. Thus, compute nodes with  prolonged wait times are more likely to successfully acquire the latch through more frequent latch retries.

\textcolor{black}{To addressed global write starvation induced by continuous global reads, the protocol employs a spin-waiting mechanism. Specifically, when a high-priority invalidation message is detected on reader nodes, a flag is set in the corresponding local cached frame, compelling subsequent global readers to spin for a predetermined duration. The spin duration of the global readers corresponds to the priority of the starved writer. The spin duration is proportional to the priority of the starved writer, thereby creating a sufficiently large time window during which no concurrent reader holds the targeted shared latch and enabling a concurrent writer to preempt latch ownership.
Although this approach addresses starvation, readers still achieve higher throughput owing to their inherently greater concurrency relative to writers. To balance performance between read and write operations, when write starvation is detected, its priority is recorded in the cache frame; consequently, this GCL cache frame must be invalidated by a global reader with equal or greater priority. By this priority matching mechanism, the performance of global readers and writers can be balanced given skewed workloads (See \fig~\ref{fig:fairness} b)   }
\textcolor{black}{Although the measurements above address the fairness issue, they also introduce additional space overhead, approximately 70 bytes per cached GCL.}


\section{SELCC as Abstraction layer}
\label{sec:application}

As cache coherence is addressed in SELCC, we observe that it can be used to build an abstraction layer to simplify building databases over disaggregated memory. This is because many existing database data structures and algorithms can be easily migrated, as the issues of RDMA access atomicity and cache coherence have already been resolved within the abstraction layer.

In this section, we first present the programming interfaces and the consistency model of the abstraction layer. Then, we use two examples, indexes and transaction engines, to demonstrate how to leverage the programming interfaces to build efficient disaggregated indexes and disaggregated transaction engines.

\subsection{Programming Interface}
SELCC exposes a straightforward interface to upper-level applications (Table~\ref{tab:APIs}).
Users can allocate or deallocate global cache lines by calling \texttt{Allocate/Free}. Each data access is conducted via the local cache, and has to be protected by an SELCC latch that consists of a hierarchical data structure containing a local latch in the cache frame and a global latch in the remote memory. 
The acquisition of a SELCC latch (\texttt{SELCC\_SLock/SELCC\_XLock}) ensures that both the local and global latch are obtained, thereby guaranteeing access atomicity and cache coherence across compute nodes. Upon acquisition, the API returns a cache handle pointing to the local copy of the target GCL. \textcolor{black}{Due to the lazy  latch-release introduced in \sect~\ref{sec:SELCCOV}}, the release of the SELCC latch (\texttt{SELCC\_SUnLock/SELCC\_XUnLock}) only ensures the immediate release of the local latch while deferring the release of the global latch until another compute node accesses the same GCL. Additionally, SELCC provides APIs for global atomic operations that can be utilized to generate global timestamps or sequential numbers.

\begin{table}[htbp!]
\vspace{-0.2cm}
\centering
\caption{The API of SELCC}
\vspace{-0.4cm}
\label{tab:APIs}
\resizebox{0.42\textwidth}{!}{%
\begin{tabular}{l|l|l|l}
\Xhline{4\arrayrulewidth}
\textbf{API}           & \textbf{Input} & \textbf{Output} & \textbf{Description}                                                         \\ 
\Xhline{4\arrayrulewidth}
\textit{Allocate/Free} & NA    & gaddr  & Allocate/ free a GCL \\ \hline
\begin{tabular}[c]{@{}l@{}}\textit{SELCC\_SLock/}\\ \textit{SELCC\_XLock}\end{tabular} &
  gaddr &
  handle &
  \begin{tabular}[c]{@{}l@{}}Acquire the shared/exclusive \\permission  of  the target GCL.\end{tabular} \\ \hline
\begin{tabular}[c]{@{}l@{}}\textit{SELLC\_SUnlock}/\\ \textit{SELCC\_XUnlock}\end{tabular} &
  handle &
  NA &
  \begin{tabular}[c]{@{}l@{}}Release the shared/exclusive\\ permission of the target GCL.\end{tabular} \\ \hline
\textit{Atomic} &
  \begin{tabular}[c]{@{}l@{}}gaddr,\\ args\end{tabular} &
  \begin{tabular}[c]{@{}l@{}}latch\\word \end{tabular}&
  \begin{tabular}[c]{@{}l@{}}Conduct RDMA atomic operation \\ on the given global address.\end{tabular} \\ \Xhline{4\arrayrulewidth}
\end{tabular}
}
\vspace{-0.2cm}
\end{table}

\subsection{Consistency Model}

SELCC guarantees the  highest level of consistency: sequential consistency. 
Every compute node should observe operations from different compute nodes in the same sequential order~\cite{Lamport79,sorin2022primer,attiya1994sequential}
Sequential consistency is essential for a general cache framework because many applications, 
e.g.,
banking and financial services, rely on strong consistency to provide reliable and accurate services to users. 
The primary reason for SELCC achieving sequential consistency 
is its latch-based design combined with synchronized invalidation messages. The latch, which is acquired before reads or writes, acts as a barrier, preventing reordering of operations within a thread.  
The invalidation messages mechanism ensures that any conflicting read or write can only proceed after the corresponding invalidation message has been processed.  This mechanism guarantees that, before a compute node modifies data in disaggregated memory, it must invalidate all existing cache copies, forcing all subsequent reads to fetch the most up-to-date data from disaggregated memory.  Consequently, when a compute node releases an SELCC exclusive latch, all other nodes can simultaneously be able to observe the write, establishing a total order of writes that is consistent across the system.

\begin{figure}[htbp]
\vspace{-0.2cm}
\centering
\includegraphics[width=0.9\columnwidth]{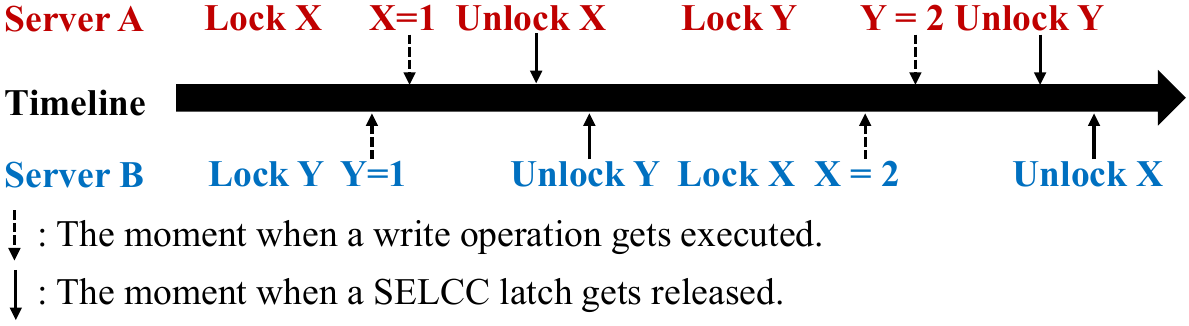}
\vspace{-0.3cm}
\caption{Example of writes from two compute nodes}
\vspace{-0.4cm}
\label{fig:MemOrder}
\end{figure}
The total order of writes can be determined by the moment the writer call \texttt{SELCC\_XUnlock}.
As shown in \fig~\ref{fig:MemOrder}, four updates occur to the disaggregated shared memory in chronological order from left to right.  The total order of the operations is ($X = 1 \rightarrow Y = 1 \rightarrow Y = 2 \rightarrow X = 2$) that is determined by the exact moments when the writer calls \texttt{SELCC\_XUnlock}. This total order may differ from the actual sequence of write operations on the timeline, which in this example is ($Y = 1 \rightarrow X = 1 \rightarrow X = 2 \rightarrow Y = 2$).  

\textcolor{black}{By default, SELCC ensures sequential consistency. However, its consistency guarantees can be relaxed to enhance performance. For example, write operations can be buffered in a local queue and returned immediately, allowing execution order of reads and writes to be relaxed. These buffered write requests can then be processed in batches, enabling the multiplexing of invalidation messages and data transfers, which reduces RDMA bandwidth consumption. By carefully setting deadlines for processing the buffered write requests, SELCC can achieve varying levels of consistency, ranging from sequential consistency to eventual consistency.}

\subsection{Index Support}

\textcolor{black}{Leveraging the abstraction layer provided by SELCC, disaggregated indexes can be implemented with ease, either from scratch or by modifying existing codebases.  For instance, we developed a concurrent B-link tree over SELCC using approximately 1,200 lines of code, of which around 200 lines related to SELCC, and modified an existing 1,400-line codebase to implement a concurrent R-tree with only 190 lines of changes. The primary steps in migrating an existing index to disaggregated memory involve aligning the index's block with the Global Cache Line (GCL) and replacing mutex-based latches with SELCC latches. Nevertheless, not all index types can be migrated with such minimal modifications. In cases where the original data structure is not completely organized into blocks (e.g., bw-tree~\cite{bwtree} and skip list~\cite{skiplistPugh90}), the data organization may require redesign; for instance, the sibling nodes in a skip list must be reorganized within the same GCL to avoid unnecessary RDMA operations during index traversal. Moreover, certain in-memory index optimizations, such as cache-conscious designs, remain effective in disaggregated environments, with the primary difference being that the CPU cache line size is replaced by the Global Cache Line size.
We evaluate the concurrent B-link-tree's performance using the YCSB benchmark  (\fig~\ref{fig:btree}), finding it achieves competitive performance compared to state-of-the-art B-trees over shared disaggregated memory.}

\subsection{Transaction Support}
We can  migrate existing concurrency control algorithms to the disaggregated architecture by leveraging the abstraction layer enabled by SELCC as  follows. (1)~Tuples should be properly organized into GCLs. (2)~Local shared-exclusive latches are replaced with \texttt{SELCC\_XLock}/\texttt{SELCC\_SLock} locks. (3)~Algorithms that require monotonic timestamps utilize \texttt{Atomic} provided by SELCC API to perform RDMA Fetch-and-Add (FAA) operations on a global timestamp generator to obtain monotonically increasing timestamps.

We implement three types of algorithms over SELCC: Two-Phase Locking with no wait strategy (2PL), Timestamp Ordering (TO), and Optimistic Concurrency Control (OCC). Tuples are organized in a heap style, meaning they are placed in GCLs in chronological order of insertion. To ensure atomicity of tuple accesses, these accesses must be protected by  \texttt{SELCC\_XLock}/ \texttt{SELCC\_SLock} locks.
For two-phase locking, SELCC latches on the GCLs are reused for locking purposes, minimizing the RDMA round trips required by the transaction concurrency control. 
Since all transactions are executed within the same compute node via RDMA,  transaction support over SELCC does not require two-phase commit protocols.
\textcolor{black}{
\subsection{Discussion}
To support a DBMS over disaggregated memory, current SELCC APIs exhibit several potential limitations, including data replication, failure recovery, and multi-tiered DB design.
Despite various proposed solutions to data replication~\cite{murat2024swarm,wang2023replicating,BehrensJBT18,LeeMCCS22}, the main challenge for SELCC memory is to incorporate these solutions without compromising performance.  Failure recovery can leverage the two-tier ARIES protocol~\cite{Disaggregation21}, which addresses compute node and memory node failures differently. For memory node failures, data replication tolerates a limited number of faults; however, extensive node failures necessitate recovery using logs and checkpoints stored in disaggregated storage.  In contrast, compute node recovery is accelerated by logs and checkpoints maintained in disaggregated memory, enabling surviving compute nodes to retrieve and replay logs based on these checkpoints.
During failure recovery, dedicated threads scan the disaggregated memory to remove stale shared or exclusive ownerships associated with the failed compute nodes.
Finally, when disaggregated memory serves as an extended buffer pool over disaggregated storage, maintaining global cache metadata (including the page address table and LRU list) imposes additional overhead on the critical path. A potential solution could involve managing these structures separately using an RPC-based protocol, thereby minimizing the extra RDMA round trips.}

\section{Experimental Evaluation}

\noindent\textbf{Platform.}  Experiments are mostly conducted on a cluster of 16 nodes in Cloudlab~\cite{CloudLab19}. The chosen instance type is c6220 that features two Xeon E5-2650v2 processors (8 cores, 
2.6GHz) and 64GB (8GB X 8) of memory per node. The cluster is interconnected using 56 Gbps Mellanox ConnectX-3 FDR Network devices. Each server runs Ubuntu 18.04.1, and the NICs are driven by Mellanox OFED-4.9-0.1.7.
The 16 servers are divided into two groups: 8 compute servers and 8 memory servers. Asymmetrical compute and memory resources are allocated on these two types of servers. The compute servers can utilize all the CPU cores but have a limited local cache (8GB by default). The memory agents on the memory servers have full access to memory but are restricted to very limited number of CPU cores (1 core by default) using the \texttt{numactl} command.

\subsection{Evaluating SELCC}
\label{sec:microbench}



\noindent\textbf{Baselines.} 
To show the efficiency of SELCC, we compare SELCC against three competitors over disaggregated memory:
(1)~\textbf{\textit{GAM}}~\cite{GAM18}, an RPC-based cache-coherence protocol designed for distributed shared memory. 
We test GAM with different consistency models: total store order consistency and sequential consistency, corresponding to GAM (TSO) and GAM (SEQ) in the figures. (2)~\textbf{\textit{ScaleStore}}~\cite{ScaleStore22}, state-of-the-art RPC-based protocol designed for distributed shared memory over distributed shared SSDs. To adapt it for disaggregated memory, we configure it differently for compute  and memory nodes. For the memory nodes, the buffer pool sizes are sufficiently large to fully store the data in memory, eliminating SSD access
.  For the compute nodes, the system no longer allocates data and instead functions as a stateless compute server. The last baseline is \textbf{\textit{SEL}}~\cite{Tobias23}, a one-sided access framework that operates without compute-side caching. While it employs the shared-exclusive latch (SEL) to ensure RDMA access atomicity, it bypasses the cache coherence problem by disabling caching, which can lead to low performance.

We found that PolarDB~\cite{PolarDBMP} and GaussDB~\cite{GaussDBMP} have recently introduced a multi-primary design based on disaggregated memory. These systems include a module to address cache coherence, but all rely on RPC-based approaches. Since they are proprietary industrial systems, directly comparing \sys{} with them may not be feasible. However, we believe comparing \sys{} with ScaleStore~\cite{ScaleStore22} is sufficient, as it also uses an RPC-based approach (with many optimizations) and is open-source.





\noindent\textbf{Benchmarks.} We test the competitors by a micro-benchmark tool~\cite{GAM18} that allows for adjustments in sharing ratios, read/write ratios, data skewness and access locality. \textcolor{black}{The accesses in the micro-benchmark directly targets at the global address of GCL, removing the necessity of disaggregated index in the experiment.}
Each compute server issues 16 Million accesses over 24 Million allocated GCLs (2KB per GCL~\footnote{\textcolor{black}{2KB GCL size reaches a good balancing between space and performance.}}, 48GB in total).
The overall throughput with different read ratios 100\% (Read only), 95\% (Read intensive), 50\% (Write intensive), 0\% (Write only) are tested. Read only workload has similar result to read intensive workload, so we omit its result in the presentation.

\subsubsection{\textbf{Evaluating the Scalability of SELCC}}


To evaluate the scalability of the SELCC protocol, we run the benchmark under a uniformly distributed workload while varying the number of compute nodes\footnote{\textcolor{black}{The total number of nodes is temporarily increased to 32 for this experiment.}}. 
We scale the number of memory nodes in proportion to the number of compute nodes. 
\textcolor{black}{Furthermore, we increased the local cache size to 16GB to more clearly reveal the overhead associated with cache invalidation.}
We compare SELCC under various sharing ratios ($sr$), following the methodology in~\cite{PolarDBMP,GAM18,Taurus23}. The sharing ratio ($sr$) indicates the percentage of allocated data accessible by all compute nodes, while the remainder is accessed privately. When the sharing ratio is zero, the system essentially operates as a sharding-based system over disaggregated memory.

\begin{figure}[htbp]
\centering
\vspace{-0.4cm}
\includegraphics[width=1.05\columnwidth]{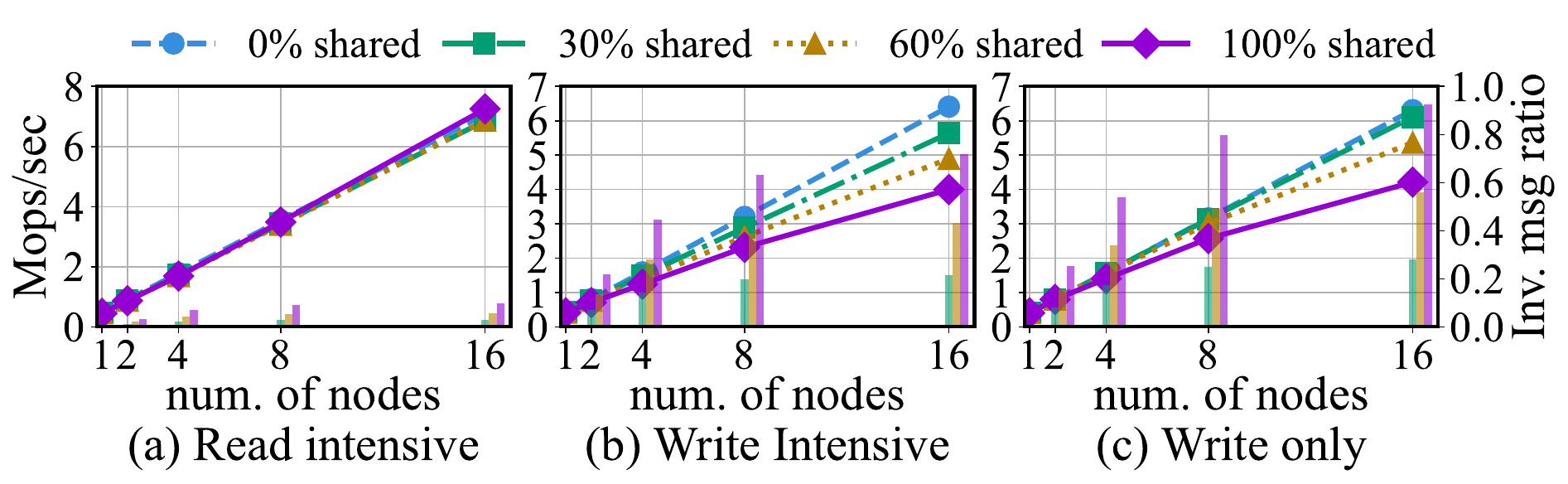}
\vspace{-0.7cm}
\caption{\textcolor{black}{Scalability of SELCC over multiple compute nodes.}}
\vspace{-0.5cm}
\label{fig:Scalability}
\end{figure}

\fig~\ref{fig:Scalability} shows the experimental results. The point values represent the overall throughput while the bar values indicate the proportion of operations requiring invalidation messages.
For the read-intensive workloads, SELCC demonstrates great scalability regardless of the sharing ratio, 
as there is very little cache coherence overhead introduced in the system. 
For write-intensive and write-only workloads, SELCC scalability  deteriorates with increased shared data ratio (\fig~\ref{fig:Scalability}c and d). The reason is that a higher shared data ratio increases the likelihood of two compute nodes caching the same data, resulting in a higher volume of invalidation messages. Compared to the fully partitioned SELCC (0\% shared ratio), the fully shared SELCC (100\% shared) shows a 37.7\%/33.2\% (8GB cache) performance degradation at 16 nodes in write-intensive and write-only workloads, respectively. 
However, SELCC still shows good scalability under write-intensive workloads. Compared to the single compute node deployment, the 16-node SELCC increases throughput by 10.4$\times$/9.83$\times$ corresponding to the write-intensive and write-only workloads.

\subsubsection{\textbf{Evaluating the Workloads with Access Locality}}
\label{sec:locality}
\begin{figure}[htbp]
\centering
\vspace{-0.4cm}
\includegraphics[width=1.00\columnwidth]{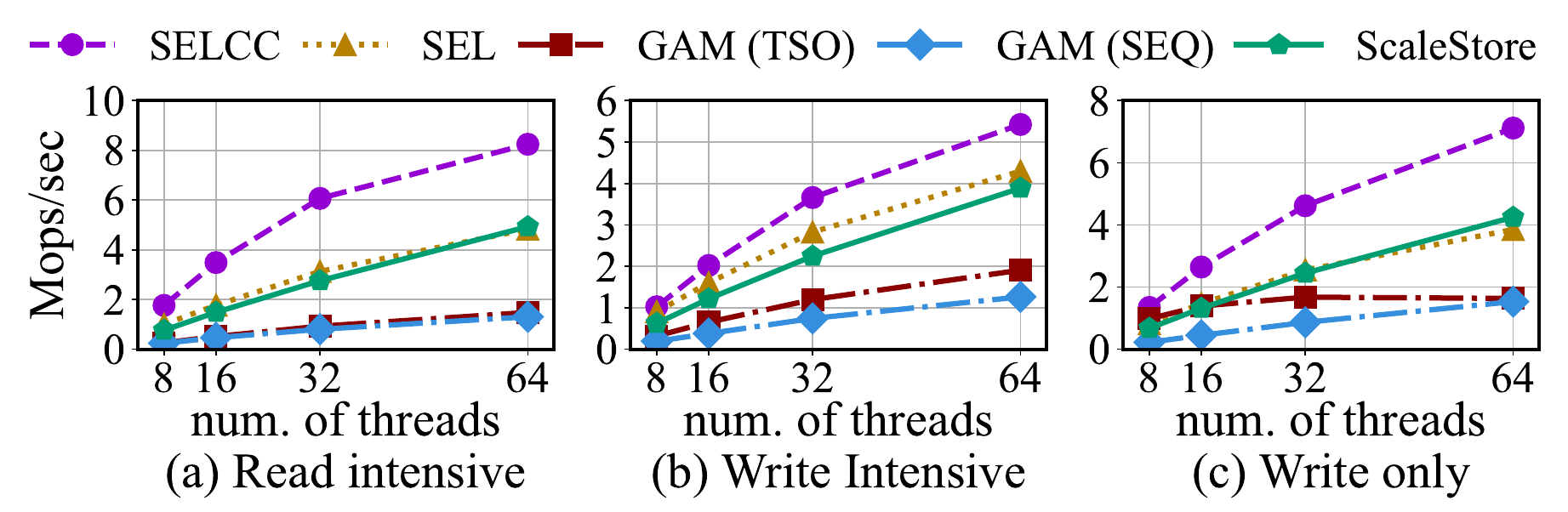}
\vspace{-0.7cm}
\caption{Performance evaluation with access locality.}
\vspace{-0.5cm}
\label{fig:locality}
\end{figure}

Workloads with access locality is one of the scenarios where the local cache can provide significant benefits.
To illustrate the performance benefits of SELCC under workloads with access locality, we adjust the uniformly distributed workload so that each operation has a 50\% probability of accessing the same GCL as the previous one.
The benchmark is executed with 8 compute nodes, 100\% sharing ratio, and varying numbers of threads across the nodes. 
Compared with SEL, due to the local cache, SELCC shows significant performance gains in all workloads (\fig~\ref{fig:locality}a), with improvements of 1.85$\times$/1.26$\times$/1.71$\times$ at 64 threads. 
Compared with GAM (TSO), GAM (SEQ) and ScaleStore, SELCC demonstrates higher performance in all four read ratios, achieving 5.57$\times$/6.31$\times$/1.67$\times$, 2.84$\times$/4.29$\times$/1.40$\times$, and 4.35$\times$/4.64$\times$/1.68$\times$ the throughput, respectively.
GAM exhibits limited thread scalability due to its serialized queue for all the read/write operations and ScaleStore's performance is limited by the insufficient remote computing power.  
\textcolor{black}{Additionally, we executed the pure-write workload on SELCC with varying GCL sizes and observed that throughput decreases as the GCL size increases (See Table below). This behavior is attributable to the fact that larger GCL sizes result in increased RDMA payloads, which in turn prolong the latency of RDMA operations.}
\begin{table}[htbp]
\vspace{-0.3cm}
\resizebox{0.8\columnwidth}{!}{%
\begin{tabular}{l|lllll}
\hline
\textbf{}                    & \textbf{512B} & \textbf{1KB} & \textbf{2KB} & \textbf{4KB} & \textbf{8KB} \\ \hline
\textbf{Throughput (Mops/s)} & 8.82          & 7.70         & 7.11         & 5.81         & 4.30         \\ \hline
\end{tabular}%
}
\vspace{-0.5cm}
\label{tab:blocksize}
\end{table}


\subsubsection{\textbf{Evaluating the Workloads with Access Skewness}}
\label{sec:skewness}

Workloads with access skewness represent another scenario where SELCC can achieve significant benefits. \textcolor{black}{Moreover, workload with extreme access skewness can trigger unfairness or even starvation of data access. 
} 

\begin{figure}[htbp]
\vspace{-0.4cm}
\centering
\includegraphics[width=1.00\columnwidth]{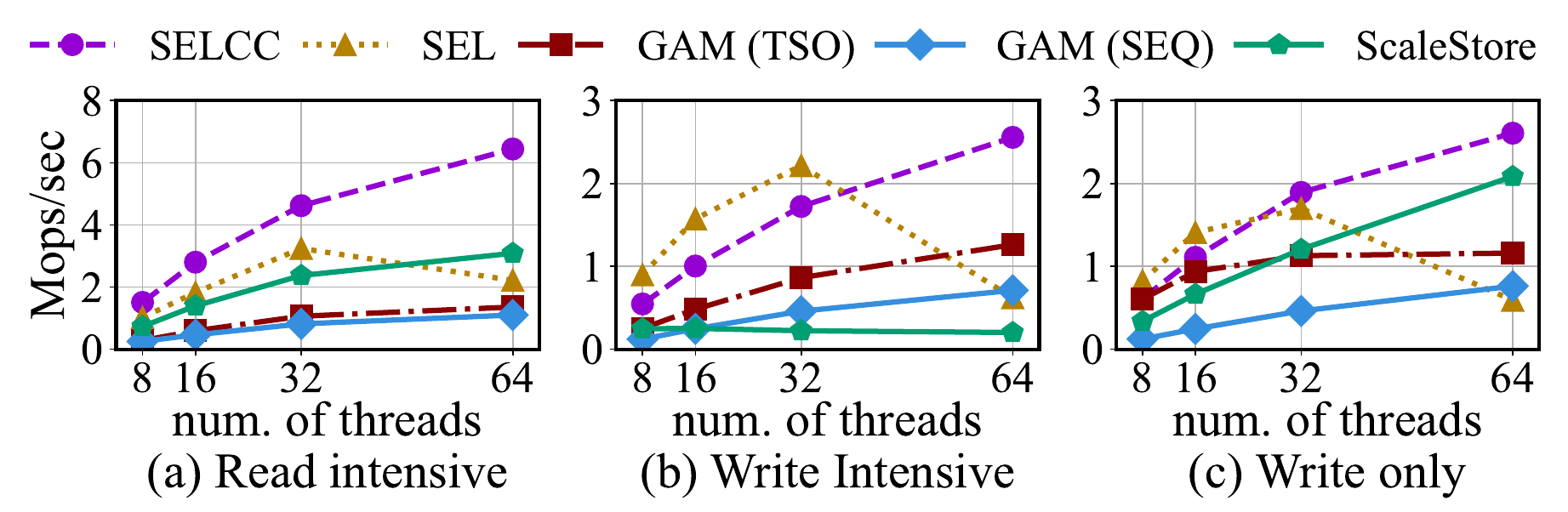}
\vspace{-0.7cm}
\caption{Performance evaluation with access skewness.}
\vspace{-0.4cm}
\label{fig:skewness}
\end{figure}

To illustrate the performance benefits of SELCC under a workload with access skewness, we run the benchmark with a Zipfian distribution, 100\% sharing ratio. The skewness parameter, $\theta$, is set to 0.99, without applying the access locality. 
Other parameters are configured in the same way as that of the previous subsection. 
For read-intensive workloads, SELCC exhibits significant performance gains, achieving throughput 3.09$\times$ over that of SEL at 64 threads. These gains result from the high cache hit ratios (60.7\%) of skewed workloads (Fig.~\ref{fig:skewness}). However, for write-intensive and write-only workloads, SEL shows better performance than SELCC initially when thread count is low, as SELCC suffers from a large number of invalidation messages triggered by the data hotspot. 
As the thread count increases, SEL experiences significant performance degradation (over 7$\times$ in write-intensive workloads),
primarily due to the high contention in RDMA atomic operations over the data hotspot. In contrast, SELCC exhibits better thread scalability due to the prioritization of local concurrency control over global concurrency control (Sec.\ref{sec:LowPrior}), shifting the bottleneck from RDMA to local write-write conflict.
Finally, SELCC outperforms GAM(TSO), GAM(SEQ) and ScaleStore by 4.74$\times$/5.85$\times$/2.09$\times$, 2.03$\times$/3.60$\times$/12.6$\times$, and 2.25$\times$/3.42$\times$/1.25$\times$ for workloads with 64 threads, highlighting the superiority of SELCC as a cache coherence protocol over disaggregated shared memory.

\begin{figure}[htbp]  
\vspace{-0.1cm}
  \centering
  \begin{tabular}{@{} c @{\hspace{0.1cm}} c @{}}
    \includegraphics[width=0.30\columnwidth]{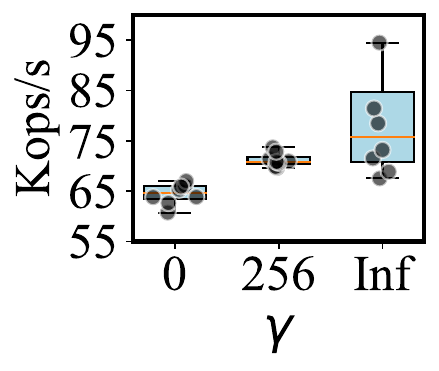} &
    \includegraphics[width=0.65\columnwidth]{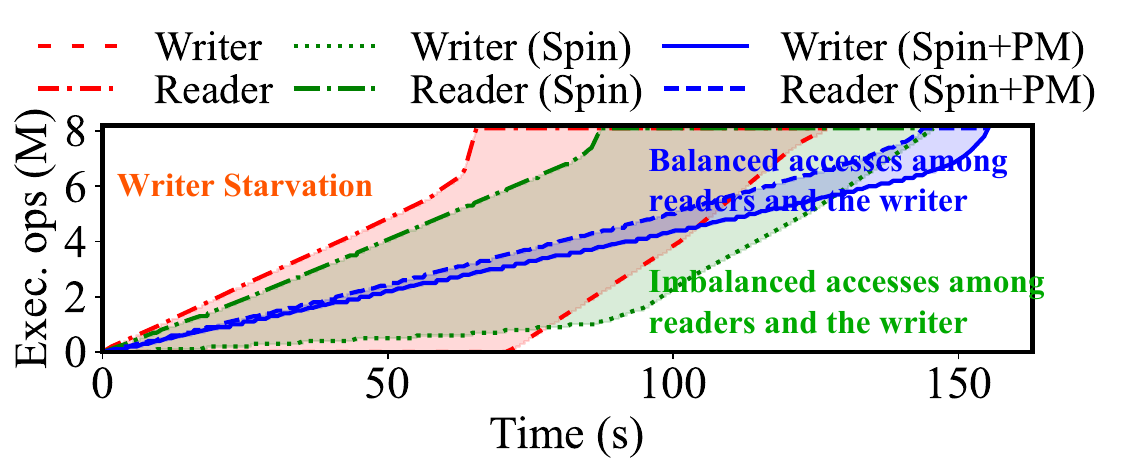} 
    \\ [-0.2cm]
    \parbox{0.35\columnwidth}{\centering \footnotesize \textbf{(a) Performance per node under write-only workload.}} &
    \parbox{0.60\columnwidth}{\centering \footnotesize \textbf{(b) Cumulative operations over time for single-writer and multiple-readers setup.}}
  \end{tabular}
  \vspace{-0.4cm}
  \caption{Performance evaluation on access fairness.}
  \vspace{-0.5cm}
  \label{fig:fairness}
\end{figure}

\textcolor{black}{In order to illustrate the effective of our approahches in improving fairness, we employ two distinct experimental setups.  First, we run a write-only workload with varying handover thresholds  $\gamma$ as introduced in \sect~\ref{sec:FairnessType1}.
Second, we configure a single-writer, multiple-readers setup, in which one compute node executes a write-only workload while the others execute read-only workloads. An ablation study of the spin-waiting and priority-matching mechanisms (detailed in Section~\ref{sec:FairnessType2}) is conducted in the setup.}
\textcolor{black}{In both experiments, we set the skewness parameter  $\theta$ to 10.
As shown in \fig~\ref{fig:fairness} a, prioritizing local ownership transfer ($\gamma=Inf$) over global ownership transfer ($\gamma=0$) improves performance but reduces access fairness. Conversely, selecting an intermediate threshold ($\gamma = 256$) effectively balances performance and fairness
\footnote{\textcolor{black}{Under highly skewed pure-write workload, the amortized latency per data access is $\frac{C T_{rm}}{\gamma} + CPT_{lm}$, where $C$ denotes the number of compute nodes, P is the thread number per compute node, $T_{rm}$ represents the latency of an RDMA roundtrip, $T_{lm}$ is the latency of main memory access.}}
. For the single-writer, multiple-readers experiment, we depict the cumulative executed operations over time for one reader node and one writer node. When no optimization is applied, the writer experiences starvation (\fig~\ref{fig:fairness} b).  After enabling spin-waiting, the writer is no longer blocked by the readers, but its throughput remains substantially lower. Once we activate the priority-matching mechanism, the writer’s performance becomes comparable to that of the readers, demonstrating significantly enhanced access fairness.}

\subsubsection{\textbf{Evaluating the Workloads with Varying Remote Computing Power}}
The benchmark described earlier is conducted on disaggregated memory with limited computing power (1 core), which makes RPC-based protocols suboptimal. 
In this subsection, we aim to demonstrate the CPU-agnostic nature of SELCC and examine whether the SELCC protocol remains effective with varying levels of remote computing power. We evaluate SELCC and ScaleStore under three different configurations: (1) Stranded remote memory, where we exhaust the computing power on memory nodes by running compute-intensive programs together with the agents on memory servers; (2) 1 remote core, the default configuration; and (3) 8 remote cores, where there is no CPU limitation and we assign eight dedicated threads for RPC processing in ScaleStore.
\begin{figure}[htbp]
\vspace{-0.4cm}
\centering
\includegraphics[width=1.00\columnwidth]{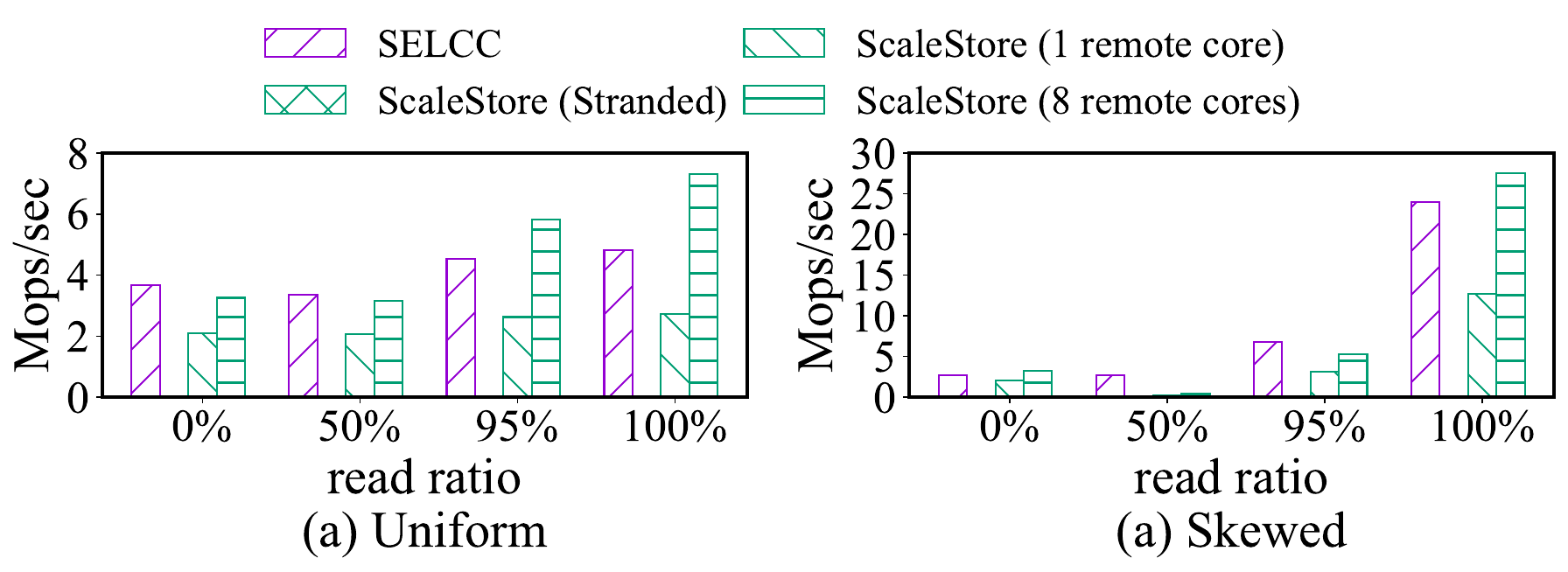}
\vspace{-0.8cm}
\caption{Performance  with varied remote computing power.}
\vspace{-0.4cm}
\label{fig:RcomputingP}
\end{figure}



As  in \fig~\ref{fig:RcomputingP}, SELCC  achieves performance comparable to ScaleStore when there is no limitation on remote CPU. SELCC slightly outperforms ScaleStore with 8 remote cores in write-intensive and pure-write workloads due to its dirty data flushing optimization (\sect~\ref{sec:Dboundaries}). The performance gap in read-intensive workloads is due to RDMA read and atomic operations reaching bandwidth limitations faster than RDMA send, receive, or write operations, as RDMA atomic and RDMA read require more processing in the RDMA NIC.
This gap could be mitigated with more advanced RDMA network cards, such as ConnectX-5.
Finally, the impact of remote computing power on the performance of RPC-based protocols is significant, especially under stranded memory conditions, where the throughput of ScaleStore is less than 0.3 Mops/s across all workloads.

\subsection{Evaluating Index Support over SELCC}
\label{sec:IndexExp}
While the previous experiments highlight the performance advantages of SELCC as a cache coherence protocol, its efficiency in supporting applications remains uncertain.
In this subsection, we construct an index following the methodology outlined in \sect~\ref{sec:application}, and evaluate its performance using 
YCSB~\cite{YCSB10}.

\noindent\textbf{Baselines.} Four B-tree baselines are evaluated in this experiment. 
The first baseline is Sherman\textcolor{black}{~\cite{Sherman2022}}, an optimized index over disaggregated shared memory. 
The second baseline is the B-tree implementation over ScaleStore~\textcolor{black}{\cite{ScaleStore22}}, configured the same as the one in the micro-benchmark. The third baseline is DEX~\textcolor{black}{\cite{lu2024dex}}, a sharding-based B-tree over disaggregated memory. Unlike the other shared-memory baselines, DEX employs a sharding mechanism to bypass the cache coherence problem. The final baseline is the B-tree over SEL.

\noindent\textbf{Benchmarks \& Configuration.} We benchmark the indexes using YCSB, following methodologies established in the existing literature \cite{lu2024dex, dLSMICDE23, Sherman2022}. Each index is loaded with 2 billion key-value records (around 40GB) and tested under varying read ratios and data skewness ($\theta = 0.99$). The experiments are conducted over 8 compute nodes, with 8 threads per node.

\subsubsection{\textbf{Results of Uniform Workloads}}
\begin{figure}[htbp]
\centering
\vspace{-0.4cm}
\includegraphics[width=1.0\columnwidth]{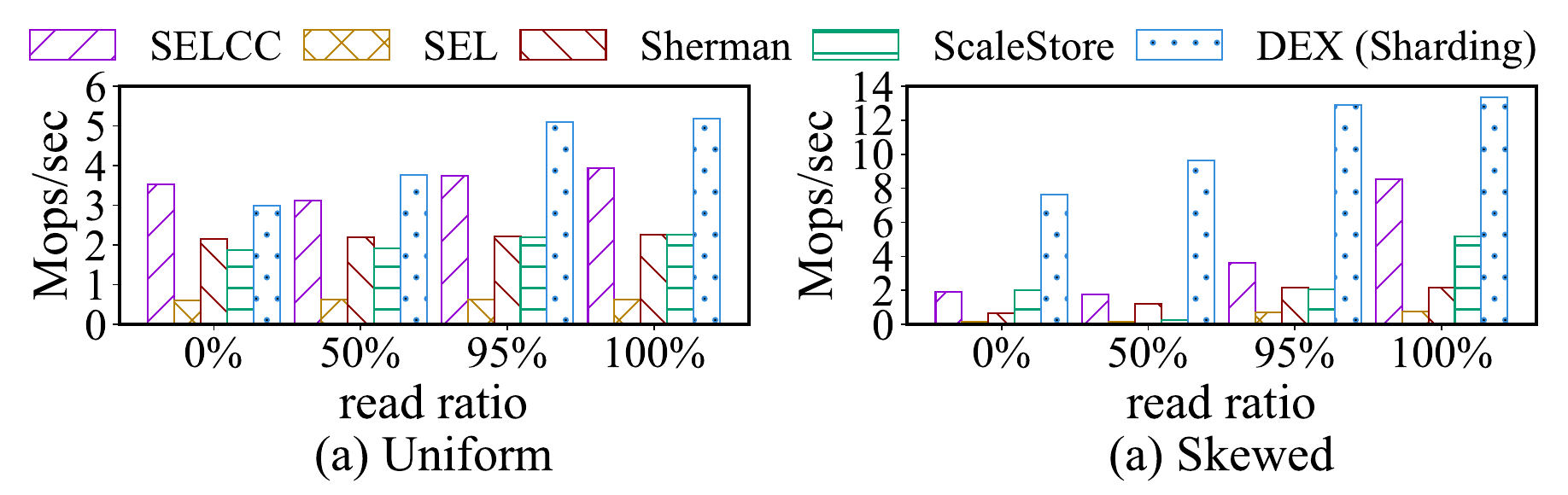}
\vspace{-0.8cm}
\caption{B-tree performance over disaggregated memory.}
\vspace{-0.5cm}
\label{fig:btree}
\end{figure}

\begin{sloppypar}
The B-tree over SELCC outperforms that over SEL by factors of 5.84$\times$/4.99$\times$/6.04$\times$/6.34$\times$, respectively (See \fig~\ref{fig:btree}a). 
Compared to Sherman, the B-tree over SELCC outperforms Sherman by factors of 1.63$\times$/1.42$\times$/1.69$\times$/1.76$\times$, respectively, because Sherman's remote synchronization requires one more RDMA round trip compared with SELCC and Sherman cannot cache leaf nodes locally. 
SELCC outperforms scalestore by factor of 1.87$\times$/1.63$\times$/1.71$\times$/1.75$\times$, respectively, showing better performance as an abstraction layer over compute-limited disaggregated memory. 
Finally, the B-tree over SELCC slightly loses to DEX under uniform workloads. This result is expected, as the sharding mechanism in DEX fully bypasses the cache coherence problem and includes many index-specific optimizations. \textcolor{black}{In contrast, the B-tree over SELCC serves as a demonstration with approximately 1,000 lines of code (LOC) over SELCC, significantly fewer than the typically extensive codebases of disaggregated indices~\cite{Sherman2022, ZuoSYZ021, RDMABtree19,dLSMICDE23,DisArt23}.
Moreover, DEX has limitations when serving as an index component in a full-fledged multi-primary database due to the overhead of cross-shard transactions (See \sect~\ref{sec:TXNExp}).} 

\end{sloppypar}

\subsubsection{\textbf{Results of Skewed workloads}}
The B-tree over SELCC outperforms both SEL and Sherman by factors of 15.3$\times$/14.6$\times$/5.03 $\times$/11.8$\times$ and 2.89$\times$/1.61$\times$/1.66$\times$/4.00$\times$, respectively, because the local cache in SELCC can hold most of the hot data. The B-tree over SEL has very limited performance under skewed workloads due to the excessive RDMA round trips required for traversing the tree. Sherman exhibits weaker performance than the B-tree over SELCC, because its leaf nodes cannot be cached locally, resulting in high RDMA atomic traffic contention over the hot spots. In contrast, SELCC can mitigate this traffic by pre-resolving conflicts in the local cache. SELCC outperform ScaleStore by 1.02$\times$/7.84$\times$/1.75$\times$/1.68$\times$ respectively, aligning with our micro-benchmark results in \sect~\ref{sec:microbench}. DEX demonstrates extremely fast performance as it completely avoids concurrency control and caches data locally. 

\subsubsection{\textbf{Ablation Study}}
We evaluate the impact of several optimizations from \sect~\ref{sec:SELCC}~and~\ref{sec:InsOpts} using the YCSB benchmark with a uniform workload and varied read ratios.
We start with 3 key optimizations disabled (GCL Forwarding in \sect~\ref{sec:PageForward}, Batched Eviction in \sect~\ref{sec:batcheviction}, and dynamic dirty boundaries in \sect~\ref{sec:Dboundaries}) and then  re-enable them one at a time. First, enabling GCL Forwarding results in substantial improvements for write-only and write-intensive workloads, with gains of 40\% and 17\%, respectively. Next, we activate the Batched Eviction, which removes eviction overhead from the critical path. This optimization yields performance improvements across all workloads (6\%, 6\%, 10\%, and 17\%, respectively). Finally, we enable the dynamic dirty boundaries optimization for GCL flushing, which significantly boosts performance for pure-write and write-intensive workloads by 11\% and 12\%, respectively.

\begin{figure}[htbp]
\centering
\vspace{-0.4cm}
\begin{tabular}{cccccc}
\includegraphics[width=\columnwidth]{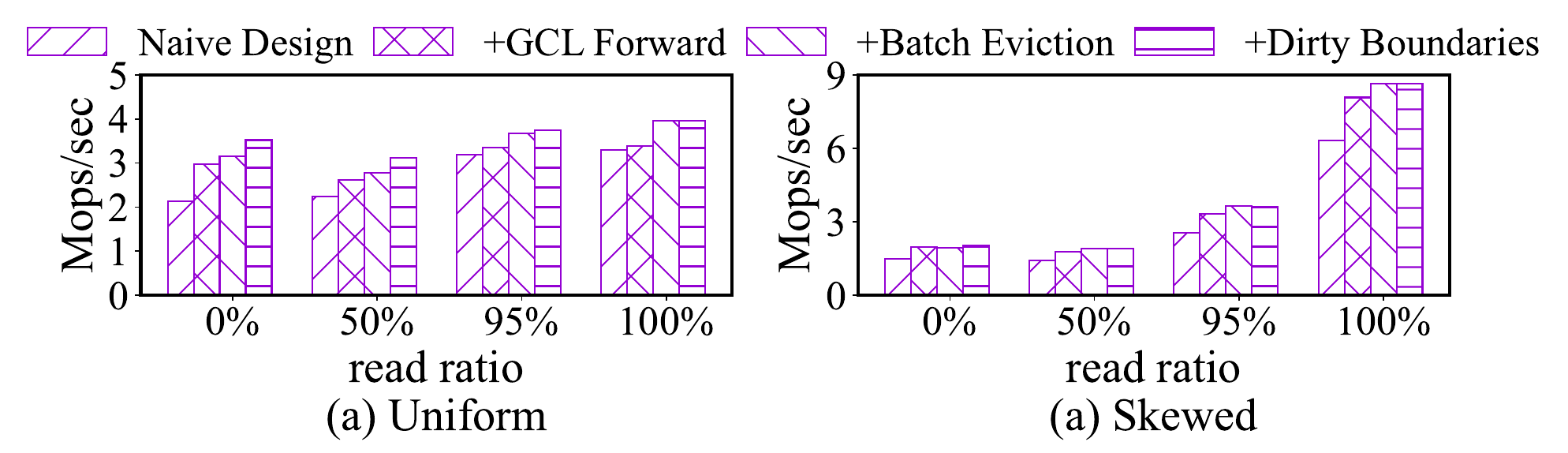}
\end{tabular}
\vspace{-0.5cm}
\caption{Ablation study}
\vspace{-0.7cm}
\label{fig:TPCC-2PC}
\end{figure}

%
\begin{figure*}[htbp]  
  \centering
  \begin{tabular}{cc}
    \includegraphics[width=0.64\textwidth]{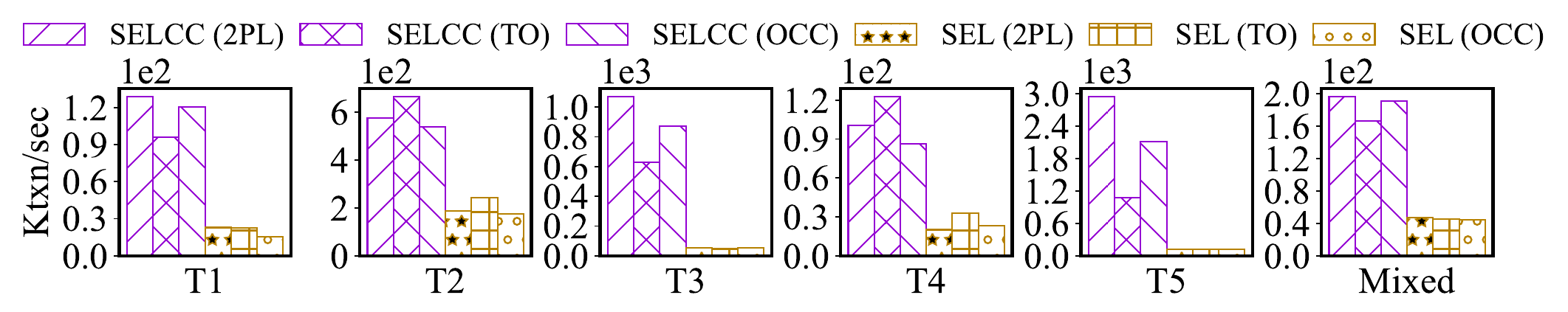} &
    \includegraphics[width=0.34\textwidth]{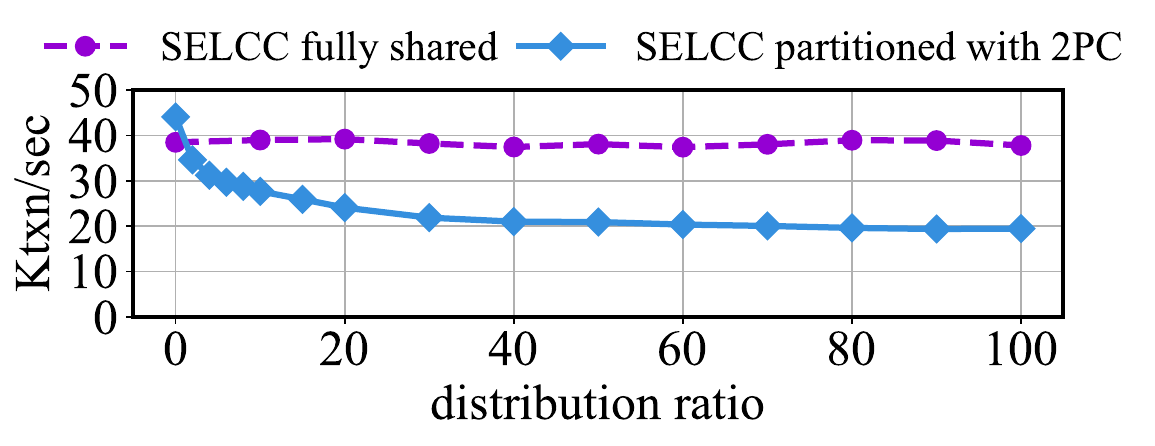} \\ [-0.2cm]
    \parbox{0.65\textwidth}{\centering \footnotesize \textbf{(a) Evaluation of different concurrency control algorithms over SELCC and SEL.}} & 
    \parbox{0.32\textwidth}{\centering \footnotesize  \textbf{(b) Fully shared vs. partitioned with 2PC.}}
  \end{tabular}
  \vspace{-0.4cm}
  \caption{TPC-C benchmark results}
  \vspace{-0.5cm}
  \label{fig:TPCC}
\end{figure*}

\subsection{Evaluating Transaction Support over SELCC}
\label{sec:TXNExp}

In this subsection, we evaluate the performance of transaction engines over SELCC using the TPC-C benchmark.

\noindent\textbf{Baselines.}
We build transactional engines using various representative concurrency control algorithms: two-phase locking (2PL) with no-wait strategy, timestamp ordering (TO), and optimistic concurrency control (OCC), following the methodology  in \sect~\ref{sec:application}. Also, we  build transaction engines over  SEL's abstraction layer to explore the benefits of cache under OLTP workloads.
Additionally, we build a 2 Phase Commit (2PC) engine over partitioned SELCC. 
By comparing the performance of fully-shared SELCC against partitioned SELCC, we aim to demonstrate the advantages of fully-shared SELCC for bypassing the two-phase commit (2PC) protocol.


\noindent\textbf{Benchmark \& Configuration.}
A database is loaded with 256 warehouses, occupying approximately 64GB of disaggregated memory. The benchmark suite includes five transactions: three of them (T1, T2 and T4) contain insertions and updates\footnote{T1: NewOrder, T2: Payment, T3: OrderStatus, T4: Delivery, T5: StockLevel~\cite{TPCC}.}. The experiment is conducted in two parts. First, we evaluate SELCC against SEL using three concurrency control algorithms, with all data fully-shared. The five transactions are first evaluated individually, and then evaluated in the standard mixed manner.
The B-tree over SELCC serves as the index for the this benchmark. Write-ahead logging is disabled to clearly highlight performance discrepancies. In the second part, we compare fully-shared SELCC against partitioned SELCC using the same database setup. The transaction concurrency control algorithm is set to 2PL, and write-ahead logging is enabled to fully demonstrate the overhead of the 2 Phase Commit protocol (2PC). 




\subsubsection{\textbf{Results for SELCC vs. SEL}}
As  in \fig~\ref{fig:TPCC}a, concurrency control algorithms over SELCC offer significant performance benefits compared to those over SEL when handling workloads generated by TPC-C. SELCC achieves up to 24.8$\times$ throughput with read-only transactions, 7.96$\times$ with update transactions, and 4.29$\times$ in mixed scenarios.
SELCC has considerable advantage over SEL even for update transactions as there are still numerous reads on immutable data (e.g., index traversal and reading immutable tables).
Also, the performance of concurrency control algorithms varies when dealing with different transactions. Algorithm TO  over SELCC exhibits poor performance in read-only transactions (T3 and T5) because even read operations require updating the read timestamp, resulting in cache invalidation. However, TO outperforms the 2PL algorithm for update transactions due to its lower abort rate.
 Generally, OCC has slower performance than 2PL as it requires acquiring the SELCC latch for the GCL twice per tuple; once during the read phase and again during the validating phase that results in a higher volume of cache invalidation messages.

\subsubsection{\textbf{Fully-Shared SELCC vs. Partitioned SELCC}} For partitioned SELCC, we partition the data according to warehouse IDs. T1 (New Order) is evaluated with varying distribution ratios, representing the percentage of cross-shard transactions. As  in \fig~\ref{fig:TPCC}b, partitioned SELCC outperforms fully-shared SELCC when the distribution ratio is 0. The gap between fully-shared and partitioned SELCC is not apparent due to slow log writing onto SSD,
shifting the bottleneck from RDMA access to SSD writes. This gap will be more significant given high-speed durable devices, e.g., persistent memory. As the number of cross-shard transactions increases, the performance of partitioned SELCC decreases significantly. This decline is attributed mainly to communication overhead and the high cost of \texttt{fsync} during both the prepare and commit stages, despite the use of group commit to reduce overhead. In contrast, the fully-shared SELCC that bypasses 2PC, remains unaffected by the distribution ratio.

\section{Related Work}
As cache coherence has been covered in Sec.~\ref{sec:background}, this section presents additional related work.

\noindent\textbf{Abstraction Layers over Distributed Shared Memory.}
Abstraction layers and unified memory models over distributed shared memory have long been  focus of research~\cite{amza1996treadmarks,CarterBZ91,LiH89,StetsDHHKPS97,GAM18,Argo15,Grappa15,FARM14,BinnigCGKZ16}.  Network latency has been a significant issue, leading many systems to install local caches with relaxed consistency models~\cite{amza1996treadmarks,CarterBZ91,LiH89,StetsDHHKPS97}. Recently, advancements in networking technologies, e.g., RDMA~\cite{GAM18,Argo15,Grappa15} and programmable switches~\cite{Concordia21}, have revitalized interest in distributed shared memory, enabling stronger consistency models for local caching. However, the interaction between local servers and remote memory relies on RPC-based communication that can saturate the limited computing resources on disaggregated memory, particularly for systems built on stranded memory.
In addition, many abstraction layers (e.g., FaRM~\cite{FARM14}, NAM~\cite{BinnigCGKZ16}) leverage one-sided RDMA as the primary method to transfer data between the local servers and the remote memory. 
However, these systems  do not apply compute-side caching to exploit data locality. 

\noindent\textbf{Database systems over disaggregated memory.}
Approaches to database research over disaggregated memory differ significantly between academia and industry. Academic research focuses on redesigning specific database components, e.g., indexes~\cite{Sherman2022,ZuoSYZ021,DisArt23,dLSMICDE23,RDMABtree19,lu2024dex} and transaction concurrency control algorithms~\cite{WeiD0C18,wang2021rdma,zamanian2016end} over the disaggregated memory. SELCC converges the individual database research by providing a layer of abstraction. 
In contrast, industry, 
e.g.,
Alibaba PolarDB and Huawei GaussDB, conducts research in  full-fledged system support over disaggregated memory~\cite{PolarDBServerless21,Disaggregation21, PolarDBMP,li2016accelerating,RuanZBMC0YLAX23}. They migrate the buffer pool onto disaggregated memory, achieving a higher cache hit ratio~\cite{Disaggregation21,PolarDBServerless21}, instant failure recovery~\cite{PolarDBServerless21,li2016accelerating}, elasticity resource provisioning~\cite{PolarDBServerless21}, and multiple primary nodes~\cite{PolarDBMP, GaussDBMP}. Existing multi-primary databases use RPC-based protocols to maintain cache coherence over disaggregated memory, but their performance could be constrained by the limited remote computing power.

\noindent\textbf{CXL-based disaggregated memory}
CXL is an emerging technology for disaggregation
~\cite{LiCXL24,AzureMD2022,CXLspec}. 
In CXL 3.0 (currently unavailable), cache coherence is expected to be guaranteed at the hardware level~\cite{CXLspec}. However, that coherence is maintained between the CPU caches and remote memory. However, this work focuses on cache coherence between the main memory in multiple compute nodes and remote memory. 
SELCC will remain valuable even with CXL 3.0, as there is still a need to cache data in the local memory of compute nodes, which introduces the cache coherence problem.





\section{Conclusion}
This paper addresses a key challenge for database systems over disaggregated memory: Maintaining cache coherence over disaggregated memory via one-sided RDMA. SELCC  provides a disaggregated memory abstraction that facilitates further research, e.g., in  indexing and transaction management. 
SELCC can be utilized by cloud-native databases to enable scalable multi-primary designs.

\section{acknowledgements}
Walid Aref acknowledges the support of the National Science Foundation under Grant Number IIS-1910216.
Jianguo Wang acknowledges the support of the National Science Foundation under Grant Number IIS-2337806.
\newpage
\bibliographystyle{ACM-Reference-Format}
\bibliography{sample}

\end{document}